\documentclass[aps,preprint,nofootinbib]{revtex4}%
\usepackage{amssymb}
\usepackage{amsmath}
\usepackage{epsfig}
\usepackage{amsfonts}
\usepackage{graphicx}
\usepackage{hyperref}%
\begin{document}
\title{Massive Superstring Scatterings in the Regge Regime}
\author{Song He}
\email{hesong@ihep.ac.cn}
\affiliation{Institute of High Energy Physics,Chinese Academic of Sciences, Beijing, PRC }
\author{Jen-Chi Lee}
\email{jcclee@cc.nctu.edu.tw}
\affiliation{Department of Electrophysics, National Chiao-Tung University and Physics
Division, National Center for Theoretical Sciences, Hsinchu, Taiwan, R.O.C. }
\author{Keijiro Takahashi}
\email{takahashi.keijiro@gmail.com}
\affiliation{Department of Electrophysics, National Chiao-Tung University, Hsinchu, Taiwan, R.O.C.}
\author{Yi Yang}
\email{yiyang@mail.nctu.edu.tw}
\affiliation{Department of Electrophysics, National Chiao-Tung University and Physics
Division, National Center for Theoretical Sciences, Hsinchu, Taiwan, R.O.C.}
\date{\today }

\begin{abstract}
We calculate four classes of high energy massive string scattering amplitudes
of fermionic string theory at arbitrary mass levels in the Regge regime (RR).
We show that all four leading order amplitudes in the RR can be expressed in
terms of the Kummer function of the second kind. Based on the summation
algorithm of a set of extended signed Stirling number identities, we show that
all four ratios calculated previously by the method of decoupling of zero-norm
states among scattering amplitudes in the Gross Regime (GR) can be extracted
from this Kummer function in the RR. Finally, we conjecture and give evidences
that the existence of these four GR ratios in the RR persists to subleading
orders in the Regge expansion of all high energy fermionic string scattering amplitudes.

\end{abstract}
\maketitle
\tableofcontents
%

\setcounter{equation}{0}
\renewcommand{\theequation}{\arabic{section}.\arabic{equation}}%

\section{Introduction}

High energy, fixed angle limit of string scattering amplitudes \cite{GM,
Gross, GrossManes} had been used to probe the fundamental spacetime symmetry
of string theory \cite{Gross}. In this approach, one needs to calculate
infinite number of massive string scattering amplitudes. By taking high energy
limit of the calculation, a lot of mathematical simplifications result and
many interesting characteristics of high energy behavior of the theory can be
obtained. There are two fundamental regimes of high energy string scattering
amplitudes. These are the fixed angle regime or Gross regime (GR), and the
fixed momentum transfer regime or Regge regime (RR). These two regimes
represent two different high energy perturbation expansions of the scattering
amplitudes, and contain complementary information of the theory. The high
energy string scattering amplitudes in the GR \cite{GM, Gross, GrossManes}
were recently intensively reinvestigated for massive string states at
arbitrary mass levels \cite{ChanLee1,ChanLee2,
CHL,CHLTY,PRL,susy,Closed,Decay,Compact,CC}. See also the developments in
\cite{West1,West2,Moore}. An infinite number of linear relations, or stringy
symmetries, among string scattering amplitudes of different string states were
obtained. Moreover, these linear relations can be solved for each fixed mass
level, and ratios $T^{(N,2m,q)}/T^{(N,0,0)}$ among the amplitudes can be
obtained. An important new ingredient of these calculations is the decoupling
of zero-norm states (ZNS) \cite{ZNS1,ZNS3,ZNS2} in the old covariant first
quantized (OCFQ) string spectrum, in particular, the identification of
inter-particle symmetries induced by the inter-particle ZNS \cite{ZNS1} in the spectrum.

Another fundamental regime of high energy string scattering amplitudes is in
the RR \cite{RR1,RR2,RR3,RR4,RR5,RR6}. See also \cite{OA,DL,KP}. An
interesting breakthrough of the subject was made in 2008 \cite{bosonic}
through the calculation of high energy string scattering amplitudes for
arbitrary mass levels in the RR. It turns out that both the saddle-point
method and the method of decoupling of high energy ZNS adopted in the
calculation of GR do not apply to the case of RR. However, a direct
calculation to get the complete form of the amplitudes is achievable and the
general formula for the high energy scattering amplitudes for each fixed mass
level in the RR can be written down explicitly. It was found that the number
of high energy scattering amplitudes for each fixed mass level in the RR is
much more numerous than that of GR calculated previously. In contrast to the
case of scatterings in the GR, there is no linear relation among scatterings
in the RR. Moreover, it was discovered that the leading order amplitudes at
each fixed mass level in the RR can be expressed in terms of the Kummer
function of the second kind. Furthermore, for those leading order high energy
amplitudes $A^{(N,2m,q)}$ in the RR with the same type of $(N,2m,q)$ as those
of GR, we can extract from them the ratios $T^{(N,2m,q)}/T^{(N,0,0)}$ in the
GR by using this Kummer function. Mathematically, the proof was based on a set
of summation algorithm for signed Stirling number identities derived by
Mkauers in 2007 \cite{MK}.

This new development of high energy behavior of string theory enables one to
express the ratios $T^{(N,2m,q)}/T^{(N,0,0)}$ (or symmetry) of string theory
in terms of Kummer function and thus may shed light on a deeper understanding
of algebraic structure of stringy symmetries. Mathematically, the realization
of the Stirling number identities by string theory brings an interesting
bridge between string theory and combinatorial theory. It is thus important to
probe the structure of more high energy string scattering amplitudes in this
context, and relate it to the Kummer function and more Stirling number
identities. \qquad

In this paper, we will calculate four classes of high energy massive string
scattering amplitudes of fermionic string theory in the Regge regime (RR). We
show that, as in the case of bosonic string, the leading order amplitudes in
the RR can be expressed in terms of the Kummer function of the second kind.
Based on the summation algorithm of a set of extended Stirling number
identities (among them, one remains to be proved mathematically), we show that
all four ratios calculated previously among scattering amplitudes in the Gross
Regime (GR) can be extracted from this Kummer function in the RR. We point out
that we failed to prove one of the Stirling number identity we used in the
text. This identity will be taken as an identity predicted by string theory.
We will also provide some numerical evidences in the appendix to support our
prediction. Hopefully a rigorous proof of it will be given in the near future.
Finally, we conjecture and give evidences that the existence of these four GR
ratios in the RR persists to all subleading orders in the Regge expansion of
all four high energy string scattering amplitudes for the even mass level with
$(N+1)=\frac{M_{2}^{2}}{2}$= odd. For the odd mass levels with $(N+1)=\frac
{M_{2}^{2}}{2}$= even, the existence of the GR ratios will be terminated and
shows up only in the first $\frac{N+1}{2}+1$ terms in the Regge expansion of
the amplitudes. This paper is organized as following. In section II, we
briefly review the previous calculation of high energy string scatterings in
the GR. In section III, we calculate four classes of fermonic string
scatterings in the Regge limit. Section IV is devoted to the extraction of the
Ratios of high energy amplitudes in the GR from the scattering amplitudes in
the RR. We also give proofs of a set of Stirling number identities we used in
the text. In section V, we calculate the subleading order amplitudes and
ratios. A conclusion is presented in section VI. The exact kinematic relations
of the Regge scatterings used in section V are collected in the appendix A. In
appendix B, we give a numerical proof of the master identity Eq.(\ref{4.3}) we
used intensively in the text.%

\setcounter{equation}{0}
\renewcommand{\theequation}{\arabic{section}.\arabic{equation}}%

\section{Review of Fixed Angle Scatterings}

In this section, we begin with a brief review of high energy string
scatterings in the fixed angle regime. That is in the kinematic regime%
\begin{equation}
s,-t\rightarrow\infty,\frac{t}{s}\approx-\sin^{2}\frac{\theta}{2}=\text{fixed
(but }\theta\neq0\text{)} \label{2.1}%
\end{equation}
where $s,t$ and $u$ are the Mandelstam variables and $\theta$ is the CM
scattering angle. It was shown \cite{CHLTY,PRL} that for the 26D open bosonic
string the only states that will survive the high-energy limit at mass level
$M_{2(B)}^{2}=2(N-1)$ are of the form%
\begin{equation}
\left\vert N,2m,q\right\rangle \equiv(\alpha_{-1}^{T})^{N-2m-2q}(\alpha
_{-1}^{L})^{2m}(\alpha_{-2}^{L})^{q}|0\rangle\label{relevant states}%
\end{equation}
where the polarizations of the 2nd particle with momentum $k_{2}$ on the
scattering plane were defined to be $e^{P}=\frac{1}{M_{2(B)}}(E_{2}%
,\mathrm{k}_{2},0)=\frac{k_{2}}{M_{2(B)}}$ as the momentum polarization,
$e^{L}=\frac{1}{M_{2(B)}}(\mathrm{k}_{2},E_{2},0)$ the longitudinal
polarization and $e^{T}=(0,0,1)$ the transverse polarization. Note that%
\begin{equation}
e^{P}=e^{L}\text{ in the GR,} \label{2.3}%
\end{equation}
and the scattering plane is defined by the spatial components of $e^{L}$ and
$e^{T}$. Polarizations perpendicular to the scattering plane are ignored
because they are kinematically suppressed for four point scatterings in the
high energy limit. One can use the saddle-point method to calculate the high
energy scattering amplitudes. For simplicity, we choose $k_{1}$, $k_{3}$ and
$k_{4}$ to be tachyons. It turns out that all scattering amplitudes at each
fixed mass level are proportional to each other, and the final result for the
ratios of high energy, fixed angle string scattering amplitude are
\cite{CHLTY,PRL}%
\begin{equation}
\frac{T^{(N,2m,q)}}{T^{(N,0,0)}}=\left(  -\frac{1}{M_{2(B)}}\right)
^{2m+q}\left(  \frac{1}{2}\right)  ^{m+q}(2m-1)!!. \label{ratios}%
\end{equation}
The precise definition of $T^{(N,2m,q)}$ is as following
\begin{equation}
T^{(N,2m,q)}=\langle V_{1}(\partial X_{2}^{T})^{N-2m-2q}(\partial X_{2}%
^{L})^{2m}(\partial^{2}X_{2}^{L})^{q}e^{ik\cdot X_{2}}V_{3}V_{4}\rangle.
\end{equation}
In the above equation, vertices $V_{1},$ $V_{3}$ and $V_{4}$ can be arbitrary
but fixed string states and their tensor indices are omitted. We use labels 1
and 2 for incoming particles and 3 and 4 for outgoing particles. In the center
of mass frame, the scattering angle $\theta$ is defined to be the angle
between $\overrightarrow{k}_{1}$ and $\overrightarrow{k}_{3}$. The ratios in
Eq.(\ref{ratios}) can also be obtained by using the decoupling of two types of
ZNS in the spectrum%
\begin{equation}
\text{Type I}:L_{-1}\left\vert x\right\rangle ,\text{ where }L_{1}\left\vert
x\right\rangle =L_{2}\left\vert x\right\rangle =0,\text{ }L_{0}\left\vert
x\right\rangle =0;
\end{equation}%
\begin{equation}
\text{Type II}:(L_{-2}+\frac{3}{2}L_{-1}^{2})\left\vert \widetilde
{x}\right\rangle ,\text{ where }L_{1}\left\vert \widetilde{x}\right\rangle
=L_{2}\left\vert \widetilde{x}\right\rangle =0,\text{ }(L_{0}+1)\left\vert
\widetilde{x}\right\rangle =0.
\end{equation}
While Type I states have zero-norm at any space-time dimension, Type II states
have zero-norm only at D=26. As examples, for $M_{2(B)}^{2}=4,6$, we get
\cite{ChanLee1,ChanLee2}
\begin{equation}
T_{TTT}:T_{LLT}:T_{(LT)}:T_{[LT]}=8:1:-1:-1, \label{CL}%
\end{equation}%
\begin{equation}%
\begin{array}
[c]{ccccccccccccccccc}%
T_{TTTT} & : & T_{TTLL} & : & T_{LLLL} & : & T_{TTL} & : & T_{LLL} & : &
\tilde{T}_{LT,T} & : & \tilde{T}_{LP,P} & : & T_{LL} & : & \tilde{T}_{LL}\\
16 & : & \frac{4}{3} & : & \frac{1}{3} & : & -\frac{4\sqrt{6}}{9} & : &
-\frac{\sqrt{6}}{9} & : & -\frac{2\sqrt{6}}{3} & : & 0 & : & \frac{2}{3} & : &
0
\end{array}
. \label{CL2}%
\end{equation}
In the above two equations, the authors of \cite{ChanLee1,ChanLee2} had used
another basis (corresponding to states listed by Young diagrams) to define the
amplitudes. For example
\begin{equation}
\mathcal{T}_{(LT)}=\langle V_{1}\left(  \partial^{2}X_{2}^{(L}\partial
X_{2}^{T)}e^{ik\cdot X_{2}}\right)  V_{3}V_{4}\rangle,\mathcal{T}%
_{[LT]}=\langle V_{1}\left(  \partial^{2}X_{2}^{[L}\partial X_{2}%
^{T]}e^{ik\cdot X_{2}}\right)  V_{3}V_{4}\rangle.
\end{equation}
These amplitudes are linear combination of $T^{(N,2m,q)\text{ }}$defined
previously. We give one specific example here. One choice of the vertex of the
spin three state at $M_{2(B)}^{2}=4$ is%

\begin{equation}
(\epsilon_{\mu\nu\lambda}\alpha_{-1}^{\mu\nu\lambda}+\epsilon_{(\mu\nu)}%
\alpha_{-1}^{\mu}\alpha_{-2}^{\nu})\left\vert 0,k\right\rangle ;\epsilon
_{(\mu\nu)}=-\frac{3}{2}k^{\lambda}\epsilon_{\mu\nu\lambda},k^{\mu}k^{\nu
}\epsilon_{\mu\nu\lambda}=0,\eta^{\mu\nu}\epsilon_{\mu\nu\lambda}=0
\end{equation}
which is conformal invariant. In the high energy limit, all components
perpendicular to the scattering plane are of subleading order in energy and
can be neglected. By using the helicity decomposition, and writing
$\epsilon_{\mu\nu\lambda}=\Sigma_{\alpha,\beta,\delta}e_{\mu}^{\alpha}e_{\nu
}^{\beta}e_{\lambda}^{\delta}u_{\alpha\beta\delta};\alpha,\beta,\delta=P,L,T,$
we can get \cite{ChanLee1,ChanLee2}%

\begin{align}
(\epsilon_{\mu\nu\lambda}\alpha_{-1}^{\mu\nu\lambda}+\epsilon_{(\mu\nu)}%
\alpha_{-1}^{\mu}\alpha_{-2}^{\nu})\left\vert 0,k\right\rangle  &
=[u_{PLT}(6\alpha_{-1}^{PLT}+6\alpha_{-1}^{(L}\alpha_{-2}^{T)})\nonumber\\
&  +u_{TTP}(3\alpha_{-1}^{TTP}-3\alpha_{-1}^{LLP}+3\alpha_{-1}^{(T}\alpha
_{-2}^{T)}-3\alpha_{-1}^{(L}\alpha_{-2}^{L)})\nonumber\\
&  +u_{TTL}(3\alpha_{-1}^{TTL}-\alpha_{-1}^{LLL})\nonumber\\
&  +u_{TTT}(\alpha_{-1}^{TTT}-3\alpha_{-1}^{LLT})]\left\vert 0,k\right\rangle
\label{PRD}%
\end{align}
where $\alpha_{-1}^{\mu\nu\lambda}\equiv\alpha_{-1}^{\mu}\alpha_{-1}^{\nu
}\alpha_{-1}^{\lambda}$ etc. It can be shown that the ratios and amplitudes
calculated in these two bases (one with Young tableaux and the other one no)
were consistent with each other. However, the calculation for general mass
levels is much easier to perform in the basis defined in
Eq.(\ref{relevant states}). Note that Eq.(\ref{PRD}) is valid also in the RR.

We now consider the fermionic string case. In this paper, we will only
consider high energy scattering amplitudes of string states with polarizations
on the scattering plane. Some high energy scatterings of string states with
polariations orthogonal to the scattering plane in the GR were discussed in
\cite{susy}. It was shown that there are four types of high energy string
scattering amplitudes for states in the NS sector with even GSO parity which
can be written down explicitly for the mass level $M_{2}^{2}=2(N+1)$\ as (here
we have replaced all $e^{L}$ in \cite{susy} by $e^{P}$ for the purpose of the
following discussion in sections III and IV)%
\begin{align}
\left\vert N+1,2m,q\right\rangle \otimes\left\vert b_{-\frac{1}{2}}%
^{T}\right\rangle  &  \equiv(\alpha_{-1}^{T})^{N-2m-2q+1}(\alpha_{-1}%
^{P})^{2m}(\alpha_{-2}^{P})^{q}(b_{-\frac{1}{2}}^{T})\left\vert
0,k\right\rangle ,\label{2.9}\\
\left\vert N+1,2m+1,q\right\rangle \otimes\left\vert b_{-\frac{1}{2}}%
^{P}\right\rangle  &  \equiv(\alpha_{-1}^{T})^{N-2m-2q}(\alpha_{-1}%
^{P})^{2m+1}(\alpha_{-2}^{P})^{q}(b_{-\frac{1}{2}}^{P})\left\vert
0,k\right\rangle ,\label{2.10}\\
\left\vert N,2m,q\right\rangle \otimes\left\vert b_{-\frac{3}{2}}%
^{P}\right\rangle  &  \equiv(\alpha_{-1}^{T})^{N-2m-2q}(\alpha_{-1}^{P}%
)^{2m}(\alpha_{-2}^{P})^{q}(b_{-\frac{3}{2}}^{P})\left\vert 0,k\right\rangle
,\label{2.11}\\
\left\vert N-1,2m,q-1\right\rangle \otimes\left\vert b_{-\frac{1}{2}}%
^{T}b_{-\frac{1}{2}}^{P}b_{-\frac{3}{2}}^{P}\right\rangle  &  \equiv
(\alpha_{-1}^{T})^{N-2m-2q}(\alpha_{-1}^{P})^{2m}(\alpha_{-2}^{P}%
)^{q-1}(b_{-\frac{1}{2}}^{T})(b_{-\frac{1}{2}}^{P})(b_{-\frac{3}{2}}%
^{P})\left\vert 0,k\right\rangle . \label{2.12}%
\end{align}
Note that the number of $\alpha_{-1}^{P}$ operator in Eq.(\ref{2.10}) is odd.
In the OCFQ spectrum of open superstring, the solutions of physical state
conditions include positive-norm propagating states and two types of zero-norm
states. In the NS sector, the latter are \cite{GSW}%
\begin{equation}
\text{Type I}:G_{-\frac{1}{2}}\left\vert x\right\rangle ,\text{ where
}G_{\frac{1}{2}}\left\vert x\right\rangle =G_{\frac{3}{2}}\left\vert
x\right\rangle =0,\text{ }L_{0}\left\vert x\right\rangle =0; \label{2.13}%
\end{equation}%
\begin{equation}
\text{Type II}:(G_{-\frac{3}{2}}+2G_{-\frac{1}{2}}L_{-1})\left\vert
\widetilde{x}\right\rangle ,\text{ where }G_{\frac{1}{2}}\left\vert
\widetilde{x}\right\rangle =G_{\frac{3}{2}}\left\vert \widetilde
{x}\right\rangle =0,\text{ }(L_{0}+1)\left\vert \widetilde{x}\right\rangle =0.
\label{2.14}%
\end{equation}
While Type I states have zero-norm at any space-time dimension, Type II states
have zero-norm only at D=10. It was shown that \cite{susy}, for each fixed
mass level, all high energy scattering amplitudes corresponding to states in
Eqs.(\ref{2.9})-(\ref{2.12}) are proportional to each other, and the ratios
can be determined from the method of decoupling of two types of zero-norm
states, Eqs.(\ref{2.13}) and (\ref{2.14}), or the method of Virasoro
constraints in the high energy limit. These ratios were calculated to be
\cite{susy}%
\begin{align}
\left\vert N,2m,q\right\rangle \otimes\left\vert b_{-\frac{3}{2}}%
^{P}\right\rangle  &  =\left(  -\frac{1}{2M_{2}}\right)  ^{q+m}\frac{\left(
2m-1\right)  !!}{\left(  -M_{2}\right)  ^{m}}\left\vert N,0,0\right\rangle
\otimes\left\vert b_{-\frac{3}{2}}^{P}\right\rangle ,\label{2.15}\\
\left\vert N+1,2m+1,q\right\rangle \otimes\left\vert b_{-\frac{1}{2}}%
^{P}\right\rangle  &  =\left(  -\frac{1}{2M_{2}}\right)  ^{q+m}\frac{\left(
2m+1\right)  !!}{\left(  -M_{2}\right)  ^{m+1}}\left\vert N,0,0\right\rangle
\otimes\left\vert b_{-\frac{3}{2}}^{P}\right\rangle ,\label{2.16}\\
\left\vert N+1,2m,q\right\rangle \otimes\left\vert b_{-\frac{1}{2}}%
^{T}\right\rangle  &  =\left(  -\frac{1}{2M_{2}}\right)  ^{q+m}\frac{\left(
2m-1\right)  !!}{\left(  -M_{2}\right)  ^{m-1}}\left\vert N,0,0\right\rangle
\otimes\left\vert b_{-\frac{3}{2}}^{P}\right\rangle ,\label{2.17}\\
\left\vert N-1,2m,q-1\right\rangle \otimes\left\vert b_{-\frac{1}{2}}%
^{T}b_{-\frac{1}{2}}^{P}b_{-\frac{3}{2}}^{P}\right\rangle  &  =\left(
-\frac{1}{2M_{2}}\right)  ^{q+m}\frac{\left(  2m-1\right)  !!}{\left(
-M_{2}\right)  ^{m}}\left\vert N,0,0\right\rangle \otimes\left\vert
b_{-\frac{3}{2}}^{P}\right\rangle . \label{2.18}%
\end{align}
Note that, in order to simplify the notation, we have only shown the second
state of the four point functions to represent the scattering amplitudes on
both sides of each equation above. This notation will be used throughout the
paper whenever is necessary. Eqs.(\ref{2.15}) to (\ref{2.18}) are thus the
SUSY generalization of Eq.(\ref{ratios}) for the bosonic string. In the next
section, in contrast to the ZNS method used in the GR, we will used a direct
calculation method to calculate the scattering amplitudes for general mass
levels in the RR. Furthermore, we can use these amplitudes to extract the
ratios in Eqs.(\ref{2.15}) to (\ref{2.18}) calculated above.%

\setcounter{equation}{0}
\renewcommand{\theequation}{\arabic{section}.\arabic{equation}}%

\section{Four Classes of Regge Scatterings}

We now turn to the discussion on high energy string scatterings in the Regge
regime. That is in the kinematic regime%
\begin{equation}
s\rightarrow\infty,\sqrt{-t}=\text{fixed (but }\sqrt{-t}\neq\infty).
\end{equation}
Instead of using $(E,\theta)$ as the two independent kinematic variables in
the GR, we choose to use $(s,t)$ in the RR. One of the reasons has been that
$t\sim E\theta$ is fixed in the RR, and it is more convenient to use $(s,t)$
rather than $(E,\theta).$ In the RR, to the lowest order, equations
(\ref{A13}) to (\ref{A18}) reduce to%
\begin{subequations}
\begin{align}
e^{P}\cdot k_{1}  &  =-\frac{1}{M_{2}}\left(  \sqrt{p^{2}+M_{1}^{2}}%
\sqrt{p^{2}+M_{2}^{2}}+p^{2}\right)  \simeq-\frac{s}{2M_{2}},\\
e^{L}\cdot k_{1}  &  =-\frac{p}{M_{2}}\left(  \sqrt{p^{2}+M_{1}^{2}}%
+\sqrt{p^{2}+M_{2}^{2}}\right)  \simeq-\frac{s}{2M_{2}},\\
e^{T}\cdot k_{1}  &  =0
\end{align}
and%
\end{subequations}
\begin{subequations}
\begin{align}
e^{P}\cdot k_{3}  &  =\frac{1}{M_{2}}\left(  \sqrt{q^{2}+M_{3}^{2}}\sqrt
{p^{2}+M_{2}^{2}}-pq\cos\theta\right)  \simeq-\frac{\tilde{t}}{2M_{2}}%
\equiv-\frac{t-M_{2}^{2}-M_{3}^{2}}{2M_{2}},\label{t-}\\
e^{L}\cdot k_{3}  &  =\frac{1}{M_{2}}\left(  p\sqrt{q^{2}+M_{3}^{2}}%
-q\sqrt{p^{2}+M_{2}^{2}}\cos\theta\right)  \simeq-\frac{\tilde{t}^{\prime}%
}{2M_{2}}\equiv-\frac{t+M_{2}^{2}-M_{3}^{2}}{2M_{2}},\label{t+}\\
e^{T}\cdot k_{3}  &  =-q\sin\phi\simeq-\sqrt{-{t}}.
\end{align}

Before we proceed to calculate the fermionic string scatterings for the
general mass levels in the RR, we first use a simple example of bosonic string
scattering \cite{bosonic} to illustrate a subtle difference between
scatterings in the GR and RR. In the mass level $M_{2(B)}^{2}=4$ $\left(
M_{1(B)}^{2}=M_{3(B)}^{2}=M_{4(B)}^{2}=-2\right)  $, one of the (conformal
invariant) high energy amplitudes in the RR is for the state $(\alpha
_{-1}^{TTT}-3\alpha_{-1}^{LLT})]\left\vert 0,k\right\rangle $. This can be
seen from the last line of Eq.(\ref{PRD}). For simplicity and for illustration
here, we will only calculate amplitude corresponding to the state $\alpha
_{-1}^{L}\alpha_{-1}^{L}\alpha_{-1}^{T}|0\rangle$. We stress that in order to
recover the conformal invariance, one needs to calculate the amplitude
corresponding to the state $(\alpha_{-1}^{TTT}-3\alpha_{-1}^{LLT})]\left\vert
0,k\right\rangle $. For the general mass levels, see the discussion on the
paragraph after Eq.(\ref{RR5}) below. The $s-t$ channel of this amplitude (the
$t-u$ channel amplitudes can be similarly discussed) can be calculated to be
\cite{bosonic} (we use $A$ to represent RR amplitudes and $T$ to represent GR
amplitudes respectively in this paper)%
\end{subequations}
\begin{align}
A^{LLT} &  =\int_{0}^{1}dx\cdot x^{k_{1}\cdot k_{2}}\left(  1-x\right)
^{k_{2}\cdot k_{3}}\cdot\left(  \frac{ie^{T}\cdot k_{1}}{x}-\frac{ie^{T}\cdot
k_{3}}{1-x}\right)  \left(  \frac{ie^{L}\cdot k_{1}}{x}-\frac{ie^{L}\cdot
k_{3}}{1-x}\right)  ^{2}\nonumber\\
&  \simeq-i\left(  \sqrt{-t}\right)  \left(  -\frac{1}{2M_{2}}\right)
^{2}\frac{\Gamma\left(  -\frac{s}{2}-1\right)  \Gamma\left(  -\frac{\tilde{t}%
}{2}-1\right)  }{\Gamma\left(  \frac{u}{2}+3\right)  }\nonumber\\
&  \cdot\left[  \left(  \frac{1}{4}t-\frac{9}{2}\right)  s^{3}+\left(
\frac{1}{4}t^{2}+\frac{7}{2}t\right)  s^{2}+\frac{\left(  t+6\right)  ^{2}}%
{2}s\right]  .
\end{align}
From the above calculation, one can easily see that the term $\sim\sqrt
{-t}t^{2}s^{2}$ is in the leading order in the GR, but is in the subleading
order in the RR. On the other hand, the terms $\sim\sqrt{-t}s^{3}$ is in the
subleading order in the GR, but is in the leading order in the RR. This
observation suggests that the high energy string scattering amplitudes in the
GR and RR contain information complementary to each other.

By Eqs.(\ref{t-}) and (\ref{t+}), it is important to note that%
\begin{equation}
e^{P}\neq e^{L}\text{ in the RR.}%
\end{equation}
This is very different from the case of GR. In the discussion of this section
and section IV of this paper, we will calculate the amplitudes for the
polarization $e^{P}$ and $e^{T}.$ For the additional $e^{L}$ amplitudes, the
results can be trivially modified. There is another important difference
between the high energy scattering amplitudes in the RR and in the GR. It was
found \cite{bosonic} for the bosonic string case that the number of high
energy scattering amplitudes for each fixed mass level in the RR is much more
numerous than that of GR. In fact, instead of states in
Eq.(\ref{relevant states}) for the GR, a class of high energy string states at
each fixed mass level $N=\sum_{n,m}np_{n}+mq_{m}$ for the RR are
\cite{bosonic}%
\begin{equation}
\left\vert p_{n},q_{m}\right\rangle =\prod_{n>0}(\alpha_{-n}^{T})^{p_{n}}%
\prod_{m>0}(\alpha_{-m}^{P})^{q_{m}}|0,k\rangle. \label{pq}%
\end{equation}

At this point, we note that there are other high energy vertex for the RR
which were not considered previously for the bosonic string case
\cite{bosonic}, namely%
\begin{equation}
\left\vert p_{n},q_{m},r_{l}\right\rangle =\prod_{n>0}(\alpha_{-n}^{T}%
)^{p_{n}}\prod_{m>0}(\alpha_{-m}^{P})^{q_{m}}\prod_{l>0}(\alpha_{-l}%
^{L})^{r_{l}}|0,k\rangle\label{RR5}%
\end{equation}
where $N=\sum_{n,m}np_{n}+mq_{m}+lr_{l}.$ However, for the purpose of
recovering the GR ratios, the vertex in Eq.(\ref{pq}) is good enough. All the
results in \cite{bosonic} including Kummer functions and Ratios etc. remain
the same if Eq.(\ref{RR5}) was used. However, in order to get the
\textit{conformal invariant} RR amplitudes, one needs to consider the most
general vertex in Eq.(\ref{RR5}). The calculation is similar to the one for
Eq.(\ref{pq}). For example, for vertex in Eq.(\ref{PRD}), one needs to
calculate, in addition to others, the amplitude corresponding to $\alpha
_{-1}^{PLT}|0\rangle\equiv\alpha_{-1}^{P}\alpha_{-1}^{L}\alpha_{-1}%
^{T}|0\rangle$ in the RR.

Now we come back to the discussion for the vertex in Eq.(\ref{pq}). It seems
that both the saddle-point method and the method of decoupling of high energy
ZNS adopted in the calculation of GR do not apply to the case of RR. However a
direct calculation is still manageable due to the following rules to simplify
the calculation for the leading order amplitudes in the RR:
\begin{align}
&  \alpha_{-n}^{T}:\quad\text{1 term (contraction of $ik_{3}\cdot X$ with
$\varepsilon_{T}\cdot\partial^{n}X$),}\\
&  \alpha_{-n}^{P}:%
\begin{cases}
n>1,\quad\text{1 term}\\
n=1\quad\text{2 terms}\text{ (contraction of $ik_{1}\cdot X$ and $ik_{3}\cdot
X$ with $\varepsilon_{L}\cdot\partial^{n}X$).}%
\end{cases}
\end{align}

For our purpose in this paper, we will only calculate four classes of
scattering amplitudes corresponding to states in Eq.(\ref{2.9}) to
Eqs.(\ref{2.12}) in the RR. There are much more high energy fermionic string
scattering amplitudes other than states we will consider in this paper. We
stress that, in addition to high energy scatterings of string states with
polariations orthogonal to the scattering plane considered previously in the
GR \cite{susy}, there are more high energy string scattering amplitudes with
more worldsheet fermionic operators $b_{-\frac{n}{2}}^{P,T}$ in the vertex.

\subsection{Amplitude $\left\vert N,2m,q\right\rangle \otimes\left\vert
b_{-\frac{3}{2}}^{P}\right\rangle $}

The first scattering amplitude we want to calculate corresponding to state in
Eq.(\ref{2.11}) is%
\begin{align}
A_{1}^{(N,2m,q)}  &  =\langle\psi_{1}^{T^{1}}e^{-\phi_{1}}e^{ik_{1}X_{1}}%
\cdot(\partial X_{2}^{T})^{N-2m-2q}(\partial X_{2}^{L})^{2m}(\partial^{2}%
X_{2}^{L})^{q}\partial\psi_{2}^{P}e^{-\phi_{2}}e^{ik_{2}X_{2}}\nonumber\\
&  \text{ \ \ \ }\cdot k_{\lambda3}\psi_{3}^{\lambda}e^{ik_{3}X_{3}}\cdot
k_{\sigma4}\psi_{4}^{\sigma}e^{ik_{4}X_{4}}\rangle\label{3.1}%
\end{align}
where we have dropped out an overall factor. In Eq.(\ref{3.1}), the first
vertex is a vector state in the $(-)$ ghost picture, and the last two states
are tachyons in the $(0)$ ghost picture. The second state is a tensor in the
$(-)$ ghost picture, so that the total superconformal ghost charges sum up to
$-2$. The $s-t$ channel of the amplitude can be calculated to be
\begin{align}
A_{1}^{(N,2m,q)}  &  =\int_{0}^{1}dx\,x^{k_{1}\cdot k_{2}}(1-x)^{k_{2}\cdot
k_{3}}\left[  \frac{e^{T}\cdot k_{3}}{1-x}\right]  ^{N-2m-2q}\label{3.2}\\
&  \cdot\left[  \frac{e^{P}\cdot k_{1}}{-x}+\frac{e^{P}\cdot k_{3}}%
{1-x}\right]  ^{2m}\left[  \frac{e^{P}\cdot k_{1}}{x^{2}}+\frac{e^{P}\cdot
k_{3}}{(1-x)^{2}}\right]  ^{q}\cdot\frac{1}{x}\label{3.3}\\
&  \cdot\left\{  \langle\psi_{1}^{T^{1}}\partial\psi_{2}^{P}\rangle\langle
\psi_{3}^{\lambda}\psi_{4}^{\sigma}\rangle-\langle\psi_{1}^{T^{1}}\psi
_{3}^{\lambda}\rangle\langle\partial\psi_{2}^{P}\psi_{4}^{\sigma}%
\rangle+\langle\psi_{1}^{T^{1}}\psi_{4}^{\sigma}\rangle\langle\partial\psi
_{2}^{P}\psi_{3}^{\lambda}\rangle\right\}  k_{\lambda3}k_{\sigma4}%
\label{3.4}\\
&  \simeq\int_{0}^{1}dx\,x^{k_{1}\cdot k_{2}}(1-x)^{k_{2}\cdot k_{3}}\left[
\frac{e^{T}\cdot k_{3}}{1-x}\right]  ^{N-2m-2q}\\
&  \cdot\left[  \frac{e^{P}\cdot k_{1}}{-x}+\frac{e^{P}\cdot k_{3}}%
{1-x}\right]  ^{2m}\left[  \frac{e^{P}\cdot k_{3}}{(1-x)^{2}}\right]
^{q}\cdot\frac{1}{x}\frac{1}{M_{2}}\left[  -\frac{(e^{T}\cdot k_{4}%
)(k_{2}\cdot k_{3})}{(1-x)^{2}}\right]  .
\end{align}
In Eq.(\ref{3.3}),$\frac{e^{P}\cdot k_{1}}{x^{2}}$ is of subleading order in
the RR and $\frac{1}{x}$ is the ghost contribution. The second term of
Eq.(\ref{3.4}) vanishes due to the $SL(2,R)$ gauge fixing $x_{1}%
=0,x_{2}=x,x_{3}=1$ and $x_{4}=\infty.$ The first term of Eq.(\ref{3.4})
vanishes due to $e^{T^{1}}\cdot e^{P^{2}}=0.$ The amplitude then reduces to%
\begin{align}
A_{1}^{(N,2m,q)}  &  \simeq\frac{\tilde{t}}{2M_{2}}(\sqrt{-{t}})^{N-2m-2q+1}%
\left(  \frac{\tilde{t}}{2M_{2}}\right)  ^{q}\int_{0}^{1}dx\,x^{k_{1}\cdot
k_{2}-1}(1-x)^{k_{2}\cdot k_{3}-N+2m-2}\nonumber\\
\cdot &  \sum_{j=0}^{2m}{\binom{2m}{j}}\left(  \frac{s}{2M_{2}x}\right)
^{j}\left(  \frac{-\tilde{t}}{2M_{2}(1-x)}\right)  ^{2m-j}\nonumber\\
&  =\frac{\tilde{t}}{2M_{2}}(\sqrt{-{t}})^{N-2m-2q+1}\left(  \frac{\tilde{t}%
}{2M_{2}}\right)  ^{2m+q}\nonumber\\
\cdot &  \sum_{j=0}^{2m}{\binom{2m}{j}}(-1)^{j}\left(  \frac{s}{\tilde{t}%
}\right)  ^{j}B\left(  k_{1}\cdot k_{2}-j,k_{2}\cdot k_{3}-N+j-1\right)  .
\end{align}
The Beta function above can be approximated in the large $s$, but fixed $t$
limit as follows
\begin{align}
&  B\left(  k_{1}\cdot k_{2}-j,k_{2}\cdot k_{3}+j-N-1\right) \nonumber\\
&  =B\left(  1-\frac{s}{2}+N-j,-\frac{1}{2}-\frac{t}{2}+j\right) \nonumber\\
&  =\frac{\Gamma(1-\frac{s}{2}+N-j)\Gamma(-\frac{1}{2}-\frac{t}{2}+j)}%
{\Gamma(\frac{u}{2}-1)}\nonumber\\
&  \approx B\left(  1-\frac{s}{2},-\frac{1}{2}-\frac{t}{2}\right)  \left(
1-\frac{s}{2}\right)  ^{N-j}\left(  \frac{u}{2}-1\right)  ^{-N}\left(
-\frac{1}{2}-\frac{t}{2}\right)  _{j}\nonumber\\
&  \approx B\left(  1-\frac{s}{2},-\frac{1}{2}-\frac{t}{2}\right)  \left(
-\frac{s}{2}\right)  ^{-j}\left(  -\frac{1}{2}-\frac{t}{2}\right)  _{j}%
\end{align}
where%
\begin{equation}
(a)_{j}=a(a+1)(a+2)...(a+j-1)
\end{equation}
is the Pochhammer symbol. The leading order amplitude in the RR can then be
written as%
\begin{align}
A_{1}^{(N,2m,q)}  &  \simeq\frac{\tilde{t}}{2M_{2}}B\left(  1-\frac{s}%
{2},-\frac{1}{2}-\frac{t}{2}\right)  \sqrt{-t}^{N-2m-2q+1}\left(  \frac
{1}{2M_{2}}\right)  ^{2m+q}\nonumber\\
\cdot &  (\tilde{t})^{2m+q}\sum_{j=0}^{2m}{\binom{2m}{j}}\left(  \frac
{2}{\tilde{t}}\right)  ^{j}\left(  -\frac{1}{2}-\frac{t}{2}\right)  _{j},
\label{A}%
\end{align}
which is UV power-law behaved as expected. The summation in Eq. (\ref{A}) can
be represented by the Kummer function of the second kind $U$ as follows,
\begin{equation}
\sum_{j=0}^{p}{\binom{p}{j}}\left(  \frac{2}{\tilde{t}}\right)  ^{j}\left(
-\frac{1}{2}-\frac{t}{2}\right)  _{j}=2^{p}(\tilde{t})^{-p}\ U\left(
-p,\frac{t}{2}-p+\frac{3}{2},\frac{\tilde{t}}{2}\right)  . \label{equality}%
\end{equation}
Finally, the amplitudes can be written as%
\begin{align}
A_{1}^{(N,2m,q)}  &  \simeq B\left(  1-\frac{s}{2},-\frac{1}{2}-\frac{t}%
{2}\right)  \sqrt{-t}^{N-2m-2q+1}\left(  \frac{1}{2M_{2}}\right)
^{2m+q+1}\nonumber\\
\cdot &  2^{2m}(\tilde{t})^{q+1}U\left(  -2m\,,\,\frac{t}{2}-2m+\frac{3}%
{2}\,,\,\frac{\tilde{t}}{2}\right)  . \label{A1}%
\end{align}
In the above, $U$ is the Kummer function of the second kind and is defined to
be%
\begin{equation}
U(a,c,x)=\frac{\pi}{\sin\pi c}\left[  \frac{M(a,c,x)}{(a-c)!(c-1)!}%
-\frac{x^{1-c}M(a+1-c,2-c,x)}{(a-1)!(1-c)!}\right]  \text{ \ }(c\neq2,3,4...)
\end{equation}
where $M(a,c,x)=\sum_{j=0}^{\infty}\frac{(a)_{j}}{(c)_{j}}\frac{x^{j}}{j!}$ is
the Kummer function of the first kind. $U$ and $M$ are the two solutions of
the Kummer Equation%
\begin{equation}
xy^{^{\prime\prime}}(x)+(c-x)y^{\prime}(x)-ay(x)=0.
\end{equation}
It is crucial to note that $c=\frac{t}{2}-2m+\frac{3}{2},$ and is not a
constant as in the usual case, so $U$ in Eq.(\ref{A1}) is not a solution of
the Kummer equation. This will make our analysis in section IV more complicated.

There are some important observations for the high energy amplitude in
Eq.(\ref{A1}). First, the amplitude gives the universal power-law behavior for
string states at \textit{all} mass levels%
\begin{equation}
A_{1}^{(N,2m,q)}\sim s^{\alpha(t)}\text{ \ (in the RR)}%
\end{equation}
where
\begin{equation}
\alpha(t)=a_{0}+\alpha^{\prime}t\text{, \ }a_{0}=\frac{1}{2}\text{ and }%
\alpha^{\prime}=1/2.
\end{equation}
This generalizes the high energy behavior of the four massless vector
amplitude in the RR to string states at arbitrary mass levels. Second, the
amplitude gives the correct intercept $a_{0}=\frac{1}{2}$ of fermionic string.
Finally, the amplitude can be used to reproduce the ratios in Eqs.(\ref{2.15})
calculated in the GR as we will see in section IV.

\subsection{Amplitude $\left\vert N+1,2m+1,q\right\rangle \otimes\left\vert
b_{-\frac{1}{2}}^{P}\right\rangle $}

Note that this is the only case with odd integer $2m+1.$ The scattering
amplitude corresponding to state in Eq.(\ref{2.10}) can be written as%
\begin{align}
A_{2}^{(N+1,2m+1,q)}  &  =\langle\psi_{1}^{T^{1}}e^{-\phi_{1}}e^{ik_{1}X_{1}%
}\cdot(\partial X_{2}^{T})^{N-2m-2q}(\partial X_{2}^{L})^{2m+1}(\partial
^{2}X_{2}^{L})^{q}\psi_{2}^{P}e^{-\phi_{2}}e^{ik_{2}X_{2}}\nonumber\\
&  \text{ \ \ \ }\cdot k_{\lambda3}\psi_{3}^{\lambda}e^{ik_{3}X_{3}}\cdot
k_{\sigma4}\psi_{4}^{\sigma}e^{ik_{4}X_{4}}\rangle
\end{align}
where we have dropped out an overall factor. The amplitude can be calculated
to be%
\begin{align}
A_{2}^{(N+1,2m+1,q)}  &  =\int_{0}^{1}dx\,x^{k_{1}\cdot k_{2}}(1-x)^{k_{2}%
\cdot k_{3}}\left[  \frac{e^{T}\cdot k_{3}}{1-x}\right]  ^{N-2m-2q}\nonumber\\
&  \cdot\left[  \frac{e^{P}\cdot k_{1}}{-x}+\frac{e^{P}\cdot k_{3}}%
{1-x}\right]  ^{2m+1}\left[  \frac{e^{P}\cdot k_{1}}{x^{2}}+\frac{e^{P}\cdot
k_{3}}{(1-x)^{2}}\right]  ^{q}\cdot\frac{1}{x}\nonumber\\
&  \cdot\left\{  \langle\psi_{1}^{T^{1}}\psi_{2}^{P}\rangle\langle\psi
_{3}^{\lambda}\psi_{4}^{\sigma}\rangle-\langle\psi_{1}^{T^{1}}\psi
_{3}^{\lambda}\rangle\langle\psi_{2}^{P}\psi_{4}^{\sigma}\rangle+\langle
\psi_{1}^{T^{1}}\psi_{4}^{\sigma}\rangle\langle\psi_{2}^{P}\psi_{3}^{\lambda
}\rangle\right\}  k_{\lambda3}k_{\sigma4}\nonumber\\
&  =\int_{0}^{1}dx\,x^{k_{1}\cdot k_{2}}(1-x)^{k_{2}\cdot k_{3}}\left[
\frac{e^{T}\cdot k_{3}}{1-x}\right]  ^{N-2m-2q}\left[  \frac{e^{P}\cdot k_{1}%
}{-x}+\frac{e^{P}\cdot k_{3}}{1-x}\right]  ^{2m+1}\nonumber\\
&  \cdot\left[  \frac{e^{P}\cdot k_{3}}{(1-x)^{2}}\right]  ^{q}\frac{1}%
{x}\frac{1}{M_{2}}\left[  (e^{T^{1}}\cdot k_{3})(k_{2}\cdot k_{4}%
)-\frac{(e^{T^{1}}\cdot k_{4})(k_{2}\cdot k_{3})}{1-x}\right] \nonumber\\
&  \simeq\left(  -1\right)  ^{N}\left[  \sqrt{-{t}}\right]  ^{N-2m-2q+1}%
\left(  -\frac{1}{2M_{2}}\right)  ^{2m+q+2}\tilde{t}^{2m+q+1}\sum_{j=0}%
^{2m+1}\binom{2m+1}{j}\left(  -\frac{s}{\tilde{t}}\right)  ^{j}\nonumber\\
\cdot &  \left[
\begin{array}
[c]{c}%
-\left(  s+t+1\right)  \int_{0}^{1}dx\,x^{k_{1}\cdot k_{2}-j-1}(1-x)^{k_{2}%
\cdot k_{3}-N+j-1}\\
+\tilde{t}\int_{0}^{1}dx\,x^{k_{1}\cdot k_{2}-j-1}(1-x)^{k_{2}\cdot
k_{3}-N+j-2}%
\end{array}
\right] \nonumber\\
&  \simeq\left[  \sqrt{-{t}}\right]  ^{N-2m-2q+1}\left(  \frac{1}{2M_{2}%
}\right)  ^{2m+q+2}\tilde{t}^{2m+q+1}\sum_{j=0}^{2m+1}\binom{2m+1}{j}\left(
-\frac{s}{\tilde{t}}\right)  ^{j}\nonumber\\
\cdot &  \left[
\begin{array}
[c]{c}%
-\left(  s+t+1\right)  B\left(  k_{1}\cdot k_{2}-j,k_{2}\cdot k_{3}-N+j\right)
\\
+\tilde{t}B\left(  k_{1}\cdot k_{2}-j,k_{2}\cdot k_{3}-N+j-1\right)
\end{array}
\right]  .
\end{align}
We then do an approximation for beta function similar to the calculation for
$A_{1}^{(N,2m,q)}$ and end up with
\begin{align}
A_{2}^{(N+1,2m+1,q)}  &  \simeq B\left(  1-\frac{s}{2},-\dfrac{1}{2}-\frac
{t}{2}\right)  \left[  \sqrt{-{t}}\right]  ^{N-2m-2q+1}\left(  \frac{1}%
{2M_{2}}\right)  ^{2m+q+2}\tilde{t}^{2m+q+1}\nonumber\\
&  \cdot\sum_{j=0}^{2m+1}\binom{2m+1}{j}\left[  \left(  1+t\right)  \left(
\frac{2}{\tilde{t}}\right)  ^{j}\left(  \dfrac{1}{2}-\frac{t}{2}\right)
_{j}-\tilde{t}\left(  \frac{2}{\tilde{t}}\right)  ^{j}\left(  -\dfrac{1}%
{2}-\frac{t}{2}\right)  _{j}\right] \nonumber\\
&  \simeq B\left(  1-\frac{s}{2},-\dfrac{1}{2}-\frac{t}{2}\right)  \left[
\sqrt{-{t}}\right]  ^{N-2m-2q+1}\left(  \frac{1}{2M_{2}}\right)
^{2m+q+2}2^{2m+1}\left(  \tilde{t}\right)  ^{q}\nonumber\\
&  \cdot\left[  (1+t)U\left(  -1-2m,\frac{t}{2}-2m-\frac{1}{2},\frac{\tilde
{t}}{2}\right)  -\tilde{t}U\left(  -1-2m,\frac{t}{2}-2m+\frac{1}{2}%
,\frac{\tilde{t}}{2}\right)  \right]  . \label{Odd}%
\end{align}
Note that there are two terms in Eq.(\ref{Odd}), and the first argument of the
$U$ function $a=-1-2m$ is odd. These differences will make the calculation of
the ratios in section IV more complicated. Finally, the amplitude gives the
universal power-law behavior for string states at \textit{all} mass levels
with the correct intercept $a_{0}=\frac{1}{2}$ of fermionic string.

\subsection{Amplitude $\left\vert N+1,2m,q\right\rangle \otimes\left\vert
b_{-\frac{1}{2}}^{T}\right\rangle $}

The third scattering amplitude corresponding to state in Eq.(\ref{2.9}) is%
\begin{align}
A_{3}^{\left(  N+1,2m,q\right)  }  &  =\langle\psi_{1}^{T^{1}}e^{-\phi_{1}%
}e^{ik_{1}X_{1}}\cdot(\partial X_{2}^{T})^{N-2m-2q+1}(\partial X_{2}^{L}%
)^{2m}(\partial^{2}X_{2}^{L})^{q}\psi_{2}^{T}e^{-\phi_{2}}e^{ik_{2}X_{2}%
}\nonumber\\
&  \text{ \ \ \ }\cdot k_{\lambda3}\psi_{3}^{\lambda}e^{ik_{3}X_{3}}\cdot
k_{\sigma4}\psi_{4}^{\sigma}e^{ik_{4}X_{4}}\rangle
\end{align}
where we have dropped out an overall factor. The scattering amplitude can be
calculated to be%
\begin{align}
A_{3}^{\left(  N+1,2m,q\right)  }  &  =\int_{0}^{1}dx\,x^{k_{1}\cdot k_{2}%
}(1-x)^{k_{2}\cdot k_{3}}\left[  \frac{e^{T}\cdot k_{3}}{1-x}\right]
^{N-2m-2q+1}\nonumber\\
&  \cdot\left[  \frac{e^{P}\cdot k_{1}}{-x}+\frac{e^{P}\cdot k_{3}}%
{1-x}\right]  ^{2m}\left[  \frac{e^{P}\cdot k_{1}}{x^{2}}+\frac{e^{P}\cdot
k_{3}}{(1-x)^{2}}\right]  ^{q}\cdot\frac{1}{x}\nonumber\\
&  \cdot\left\{  \langle\psi_{1}^{T^{1}}\psi_{2}^{T}\rangle\langle\psi
_{3}^{\lambda}\psi_{4}^{\sigma}\rangle-\langle\psi_{1}^{T^{1}}\psi
_{3}^{\lambda}\rangle\langle\psi_{2}^{T}\psi_{4}^{\sigma}\rangle+\langle
\psi_{1}^{T^{1}}\psi_{4}^{\sigma}\rangle\langle\psi_{2}^{T}\psi_{3}^{\lambda
}\rangle\right\}  k_{\lambda3}k_{\sigma4}\nonumber\\
&  =\int_{0}^{1}dx\,x^{k_{1}\cdot k_{2}}(1-x)^{k_{2}\cdot k_{3}}\left[
\frac{e^{T}\cdot k_{3}}{1-x}\right]  ^{N-2m-2q+1}\left[  \frac{e^{P}\cdot
k_{1}}{-x}+\frac{e^{P}\cdot k_{3}}{1-x}\right]  ^{2m}\left[  \frac{e^{P}\cdot
k_{3}}{(1-x)^{2}}\right]  ^{q}\nonumber\\
&  \cdot\frac{1}{x}\left[  \frac{(e^{T^{1}}\cdot e^{T})(k_{3}\cdot k_{4})}%
{-x}+(e^{T^{1}}\cdot k_{3})(e^{T}\cdot k_{4})-\frac{(e^{T^{1}}\cdot
k_{4})(e^{T}\cdot k_{3})}{1-x}\right] \nonumber\\
&  \simeq\left[  \sqrt{-{t}}\right]  ^{N-2m-2q+1}\left(  \frac{1}{2M_{2}%
}\right)  ^{2m+q}\tilde{t}^{2m+q}\sum_{j=0}^{2m}\binom{2m}{j}\left(  -\frac
{s}{\tilde{t}}\right)  ^{j}\nonumber\\
&  \cdot\left[  -\frac{s}{2}\int_{0}^{1}dx\,x^{k_{1}\cdot k_{2}-j-2}%
(1-x)^{k_{2}\cdot k_{3}-N+j-1}+t\int_{0}^{1}dx\,x^{k_{1}\cdot k_{2}%
-j}(1-x)^{k_{2}\cdot k_{3}-N+j-2}\right] \nonumber\\
&  \simeq\left[  \sqrt{-{t}}\right]  ^{N-2m-2q+1}\left(  \frac{1}{2M_{2}%
}\right)  ^{2m+q}\tilde{t}^{2m+q}\sum_{j=0}^{2m}\binom{2m}{j}\left(  -\frac
{s}{\tilde{t}}\right)  ^{j}\nonumber\\
&  \cdot\left[
\begin{array}
[c]{c}%
-\frac{s}{2}B\left(  k_{1}\cdot k_{2}-j-1,k_{2}\cdot k_{3}-N+j\right) \\
+tB\left(  k_{1}\cdot k_{2}-j+1,k_{2}\cdot k_{3}-N+j-1\right)
\end{array}
\right]  .
\end{align}
We then do an approximation for beta function similar to the calculation for
$A_{1}^{(N,2m,q)}$ and end up with%

\begin{align}
&  \simeq-B\left(  1-\frac{s}{2},-\dfrac{1}{2}-\frac{t}{2}\right)  \left[
\sqrt{-{t}}\right]  ^{N-2m-2q+1}\left(  \frac{1}{2M_{2}}\right)  ^{2m+q}%
\tilde{t}^{2m+q}\nonumber\\
&  \cdot\sum_{j=0}^{2m}\binom{2m}{j}\left[  \dfrac{\left(  1+t\right)  }%
{2}\left(  \frac{2}{\tilde{t}}\right)  ^{j}\left(  \dfrac{1}{2}-\frac{t}%
{2}\right)  _{j}-t\left(  \frac{2}{\tilde{t}}\right)  ^{j}\left(  -\dfrac
{1}{2}-\frac{t}{2}\right)  _{j}\right] \nonumber\\
&  \simeq-B\left(  1-\frac{s}{2},-\dfrac{1}{2}-\frac{t}{2}\right)  \left[
\sqrt{-{t}}\right]  ^{N-2m-2q+1}\left(  \frac{1}{2M_{2}}\right)
^{2m+q}2^{2m-1}\tilde{t}^{q}\nonumber\\
&  \cdot\left[  \left(  1+t\right)  U\left(  -2m,\frac{t}{2}-2m+\frac{1}%
{2},\frac{\tilde{t}}{2}\right)  -2tU\left(  -2m,\frac{t}{2}-2m+\frac{3}%
{2},\frac{\tilde{t}}{2}\right)  \right]  .
\end{align}
In this case there are again two terms as in the amplitude $A_{2}$ but with an
even argument $a=-2m$. Finally, the amplitude gives the universal power-law
behavior for string states at \textit{all} mass levels with the correct
intercept $a_{0}=\frac{1}{2}$ of fermionic string.

\subsection{Amplitude $\left\vert N-1,2m,q-1\right\rangle \otimes\left\vert
b_{-\frac{1}{2}}^{T}b_{-\frac{1}{2}}^{P}b_{-\frac{3}{2}}^{P}\right\rangle $}

The fourth scattering amplitude corresponding to state in Eq.(\ref{2.12}) is%
\begin{align}
A_{4}^{\left(  N-1,2m,q-1\right)  }  &  =\langle\psi_{1}^{T^{1}}e^{-\phi_{1}%
}e^{ik_{1}X_{1}}\cdot(\partial X_{2}^{T})^{N-2m-2q}(\partial X_{2}^{L}%
)^{2m}(\partial^{2}X_{2}^{L})^{q-1}\psi_{2}^{T}\psi_{2}^{P}\partial\psi
_{2}^{P}e^{-\phi_{2}}e^{ik_{2}X_{2}}\nonumber\\
&  \text{ \ \ \ }\cdot k_{\lambda3}\psi_{3}^{\lambda}e^{ik_{3}X_{3}}\cdot
k_{\sigma4}\psi_{4}^{\sigma}e^{ik_{4}X_{4}}\rangle
\end{align}
where we have dropped out an overall factor. The scattering amplitude can be
calculated to be%

\begin{align}
A_{4}^{\left(  N-1,2m,q-1\right)  }  &  =\int_{0}^{1}dx\,x^{k_{1}\cdot k_{2}%
}(1-x)^{k_{2}\cdot k_{3}}\left[  \frac{e^{T}\cdot k_{3}}{1-x}\right]
^{N-2m-2q}\nonumber\\
&  \cdot\left[  \frac{e^{P}\cdot k_{1}}{-x}+\frac{e^{P}\cdot k_{3}}%
{1-x}\right]  ^{2m}\left[  \frac{e^{P}\cdot k_{1}}{x^{2}}+\frac{e^{P}\cdot
k_{3}}{(1-x)^{2}}\right]  ^{q-1}\frac{1}{x}\nonumber\\
&  \cdot\langle\psi_{1}^{T^{1}}\psi_{2}^{T}\rangle\langle\psi_{2}^{P}\psi
_{4}^{\sigma}\rangle\langle\partial\psi_{2}^{P}\psi_{3}^{\lambda}\rangle
k_{\lambda3}k_{\sigma4}\nonumber\\
&  \simeq\int_{0}^{1}dx\,x^{k_{1}\cdot k_{2}}(1-x)^{k_{2}\cdot k_{3}}\left[
\frac{e^{T}\cdot k_{3}}{1-x}\right]  ^{N-2m-2q}\left[  \frac{e^{P}\cdot k_{1}%
}{-x}+\frac{e^{P}\cdot k_{3}}{1-x}\right]  ^{2m}\nonumber\\
&  \cdot\left[  \frac{e^{P}\cdot k_{3}}{(1-x)^{2}}\right]  ^{q-1}\frac{1}%
{x}\frac{1}{M_{2}^{2}}\left[  \frac{(e^{T^{1}}\cdot e^{T})(k_{2}\cdot
k_{4})(k_{2}\cdot k_{3})}{(1-x)^{2}}\right] \nonumber\\
&  \simeq\left[  \sqrt{-{t}}\right]  ^{N-2m-2q}\left(  \frac{1}{2M_{2}%
}\right)  ^{2m+q+1}\tilde{t}^{2m+q}s\nonumber\\
&  \cdot\sum_{j=0}^{2m}\binom{2m}{j}\left(  -\frac{s}{\tilde{t}}\right)
^{j}\int_{0}^{1}dx\,x^{k_{1}\cdot k_{2}-j}(1-x)^{k_{2}\cdot k_{3}%
-N+j-1}\nonumber\\
&  \simeq\left[  \sqrt{-{t}}\right]  ^{N-2m-2q}\left(  \frac{1}{2M_{2}%
}\right)  ^{2m+q+1}\tilde{t}^{2m+q}s\nonumber\\
&  \cdot\sum_{j=0}^{2m}\binom{2m}{j}\left(  -\frac{s}{\tilde{t}}\right)
^{j}B\left(  k_{1}\cdot k_{2}-j+1,k_{2}\cdot k_{3}-N+j\right)  .
\end{align}
With a similar approximation for the beta function, we get%
\begin{align}
A_{4}^{\left(  N-1,2m,q-1\right)  }  &  \simeq B\left(  1-\frac{s}{2}%
,-\dfrac{1}{2}-\frac{t}{2}\right)  \left[  \sqrt{-{t}}\right]  ^{N-2m-2q}%
\left(  \frac{1}{2M_{2}}\right)  ^{2m+q+1}\tilde{t}^{2m+q}\nonumber\\
&  \cdot\left(  1+t\right)  \sum_{j=0}^{2m}\binom{2m}{j}\left(  -\frac
{2}{\tilde{t}}\right)  ^{j}\left(  \dfrac{1}{2}-\frac{t}{2}\right)
_{j}\nonumber\\
&  =B\left(  1-\frac{s}{2},-\dfrac{1}{2}-\frac{t}{2}\right)  \left[
\sqrt{-{t}}\right]  ^{N-2m-2q}\left(  \frac{1}{2M_{2}}\right)  ^{2m+q+1}%
\nonumber\\
&  \cdot2^{2m}(\tilde{t})^{q}\left(  1+t\right)  U\left(  -2m,\frac{t}%
{2}-2m+\frac{1}{2},\frac{\tilde{t}}{2}\right)  .
\end{align}
Again the amplitude gives the universal power-law behavior for string states
at \textit{all} mass levels with the correct intercept $a_{0}=\frac{1}{2}$ of
fermionic string. In the next section we are going to use the four amplitudes
calculated in this section to extract ratios of Eqs.(\ref{2.15}) to
(\ref{2.18}) calculated in the fixed angle regime.%

\setcounter{equation}{0}
\renewcommand{\theequation}{\arabic{section}.\arabic{equation}}%

\section{Reproducing the GR ratios from the RR}

In the bosonic string calculation \cite{bosonic}, we learned that the relative
coefficients of the highest power $t$ terms in the leading order amplitudes in
the RR can be used to reproduce the ratios of the amplitudes in the GR for
each fixed mass level. Here we present an explicit example. An explicit
calculation of the high energy string scattering amplitudes to some subleading
orders in the RR for $M_{2}^{2}=4$ are
\begin{equation}
A_{TTT}\sim\frac{1}{8}\sqrt{-t}ts^{3}+\frac{3}{16}\sqrt{-t}t(t+6)s^{2}%
+\frac{3t^{3}+84t^{2}-68t-864}{64}\sqrt{-t}\,s+O(1), \label{22-4}%
\end{equation}%
\begin{align}
A_{LLT}  &  \sim\frac{1}{64}\sqrt{-t}(t-6)s^{3}+\frac{3}{128}\sqrt{-t}%
(t^{2}-20t-12)s^{2}\nonumber\\
&  \quad\quad+\frac{3t^{3}-342t^{2}-92t+5016+1728(-t)^{-1/2}}{512}\sqrt
{-t}\,s+O(1), \label{23-4}%
\end{align}%
\begin{align}
A_{(LT)}  &  \sim-\frac{1}{64}\sqrt{-t}(t+10)s^{3}-\frac{1}{128}\sqrt
{-t}(3t^{2}+52t+60)s^{2}\nonumber\\
&  \quad\quad-\frac{3[t^{3}+30t^{2}+76t-1080-960(-t)^{-1/2}]}{512}\sqrt
{-t}\,s+O(1), \label{24-4}%
\end{align}%
\begin{align}
A_{[LT]}  &  \sim-\frac{1}{64}\sqrt{-t}(t+2)s^{3}-\frac{3}{128}\sqrt
{-t}(t+2)^{2}s^{2}\nonumber\\
&  \quad\quad-\frac{(3t-8)(t+6)^{2}[1-2(-t)^{-1/2}]}{512}\sqrt{-t}\,s+O(1).
\label{25-4}%
\end{align}
We have ignored an overall irrelevant factors in the above amplitudes. Note
that the calculation of Eq.(\ref{24-4}) and Eq.(\ref{25-4}) involves amplitude
of the state $(\alpha_{-2}^{T})(\alpha_{-1}^{L})\left\vert 0,k_{2}%
\right\rangle $ which can be shown to be of leading order in the RR
\cite{bosonic}, but is of subleading order in the GR as it is not in the form
of Eq.(\ref{relevant states}). However, the contribution of the amplitude
calculated from this state will not affect the ratios $8:1:-1:-1$ in the RR
\cite{bosonic}. One can now easily see that the ratios of the coefficients of
the highest power of $t$ in these \textit{leading order} $(s^{3})$ coefficient
functions $\frac{1}{8}:\frac{1}{64}:-\frac{1}{64}:-\frac{1}{64}$ in the RR
agree with the ratios in the GR calculated in Eq.(\ref{CL}) as expected.
Moreover, one further obeservation is that these ratios remain the same for
the coefficients of the highest power of $t$ in the \textit{subleading orders}
$(s^{2})$ $\frac{3}{16}:\frac{3}{128}:-\frac{3}{128}:-\frac{3}{128}$ and $(s)$
$\frac{3}{64}:\frac{3}{512}:-\frac{3}{512}:-\frac{3}{512}$. More examples can
be found in \cite{bosonic}.

In this section, we are going to generalize the calculation to four classes of
fermionic string states for arbitrary mass levels. We will first calculate the
leading order results in this section and postpone the subleading order
calculation to the next section. We begin with the first amplitude of
Eq.(\ref{A1}).

\subsection{Ratios for $\left\vert N,2m,q\right\rangle \otimes\left\vert
b_{-\frac{3}{2}}^{P}\right\rangle $}

It is important to note that there are no linear relations among high energy
string scattering amplitudes, Eq.(\ref{A1}), of different string states for
each fixed mass level in the RR. In other words, the ratios $A_{1}%
^{(N,2m,q)}/A_{1}^{(N,0,0)}$ are $t$-dependent functions and can be calculated
to be
\begin{align}
\frac{A_{1}^{(N,2m,q)}}{A_{1}^{(N,0,0)}}  &  =\left(  -\frac{1}{2M_{2}%
}\right)  ^{2m+q}(-)^{m}(\tilde{t}+2N+1)^{-m-q}(\tilde{t})^{2m+q}\nonumber\\
&  \cdot\sum_{j=0}^{2m}(-2m)_{j}\left(  -N-1-\frac{\tilde{t}}{2}\right)
_{j}\frac{(-2/\tilde{t})^{j}}{j!} \label{4.1}%
\end{align}
where we have used Eq.(\ref{t-}) to replace $t$ by $\tilde{t}$. If the leading
order coefficients in Eq.(\ref{4.1}) extracted from the amplitudes in the RR
are to be identified with the ratios calculated in the GR in Eq.(\ref{2.15}),
we need the following identity
\begin{align}
&  \sum_{j=0}^{2m}(-2m)_{j}\left(  -L-\frac{\tilde{t}}{2}\right)  _{j}%
\frac{(-2/\tilde{t})^{j}}{j!}\label{4.2}\\
&  =0(-\tilde{t})^{0}+0(-\tilde{t})^{-1}+...+0(-\tilde{t})^{-m+1}+\frac
{(2m)!}{m!}(-\tilde{t})^{-m}+\mathit{O}\left\{  \left(  \frac{1}{\tilde{t}%
}\right)  ^{m+1}\right\}  \label{4.3}%
\end{align}
where $L=N+1$ and is an integer. The coefficients of the terms $\mathit{O}%
\left\{  \left(  1/\tilde{t}\right)  ^{m+1}\right\}  $ in Eq.(\ref{4.3}) is
irrelevant for string amplitudes. If the identity of Eq.(\ref{4.3}) obtained
from superstring theory calculation is correct, this implies that the value of
$L$ effects only in the subleading order terms $\mathit{O}\left\{  \left(
1/\tilde{t}\right)  ^{m+1}\right\}  $ in Eq.(\ref{4.3}). We will show that
this is indeed the case mathematically and numerically. In fact, we will show
numerically that the identity is valid for arbitrary \textit{real} $L$.

We will first show the cases of $L=0,1$, and then try to generalize the proof
to arbitrary integers $L$. For $L=1,$ we rewrite the nontrivial leading term
of the above summation in Eq.(\ref{4.2}) as (we have replaced $\tilde{t}$ by
$t$ here for simplicity)
\begin{align}
&  \sum_{j=0}^{2m}(-2m)_{j}\left(  -1-\frac{t}{2}\right)  \left(  -\frac{t}%
{2}\right)  _{j-1}\left(  -\frac{2}{t}\right)  ^{j}\frac{1}{j!}\nonumber\\
&  =\sum_{j=0}^{2m}(-2m)_{j}\left(  -1-\frac{t}{2}\right)  \sum_{k=0}%
^{j-1}(-1)^{j-1-k}s(j-1,k)\left(  -\frac{t}{2}\right)  ^{k}\left(  -\frac
{2}{t}\right)  ^{j}\frac{1}{j!}\nonumber\\
&  \Longrightarrow\left[  \sum_{j=m}^{2m}(-2m)_{j}s(j-1,j-m)\frac{2^{m}}%
{j!}+\sum_{j=m+1}^{2m}(-2m)_{j}s(j-1,j-m-1)\frac{2^{m}}{j!}\right]
(-t)^{-m}\nonumber\\
&  =2^{m}\sum_{j=0}^{m}(-1)^{j+m}{\binom{2m}{j+m}}s(j+m-1,j)(-t)^{-m}%
\nonumber\\
&  +2^{m}\sum_{j=1}^{m}(-1)^{j+m}{\binom{2m}{j+m}}s(j+m-1,j-1)(-t)^{-m}%
\end{align}
where we have used the signed Stirling number of the first kind $s(n,k)$ to
expand the Pochhammer symbol. The definition of $s(n,k)$ is
\begin{equation}
(x)_{n}=\sum_{k=0}^{n}(-1)^{n-k}s(n,k)x^{k}.
\end{equation}
Thus the nontrivial leading order identity of Eq.(\ref{4.3}) can be written as
($m\geqslant0$)
\begin{equation}
F(m)\equiv\sum_{j=0}^{m}(-1)^{j}{\binom{2m}{j+m}}\left[
s(j+m-1,j-1)+s(j+m-1,j)\right]  =(2m-1)!! \label{lead}%
\end{equation}
where we have used the convention that
\begin{equation}
s(m-1,-1)%
\begin{cases}
=0\quad,\text{ for $m\geqslant1$}\\
=1\quad,\text{ for $m=0$}%
\end{cases}
\quad,s(-1,0)=0,
\end{equation}
and $(2m-1)!!=0$ for $m=0.$ To apply the algorithm developed by Mkauers in
2007 \cite{MK}, we need to introduce an auxiliary variable $u$ and define
\begin{align}
F(u,m)  &  \equiv\sum_{j=0}^{m+u}(-1)^{j}{\binom{2m+u}{j+m}}\left[
s(j+m-1,j-1)+s(j+m-1,j)\right] \nonumber\\
&  \equiv f_{1}(u,m)+f_{2}(u,m)
\end{align}
where $f_{1}$ and $f_{2}$ are the two summations, each with one Stirling
number, and $F(0,m)=F(m)$. By the algorithm, both $f_{1}$, $f_{2}$ satisfy the
following recurrence relation \cite{MK}
\begin{equation}
-(1+2m+u)f(u,m)+(2m+u)f(u+1,m)+f(u,m+1)=0, \label{recurrence}%
\end{equation}
hence, so is $F.$ Eq.(\ref{recurrence}) is the most nontrivial step to prove
Eq.(\ref{lead}). Now, note that
\begin{equation}
F(u,0)=\sum_{j=0}^{u}(-1)^{j}{\binom{u}{j}}=%
\begin{cases}
1\quad,u=0\\
0\quad,u>0
\end{cases}
.
\end{equation}
Using the recurrence relation Eq.(\ref{recurrence}) and substituting
$(u,m)=(1,0),(2,0)\cdots$, one can prove that
\begin{equation}
F(u,1)=0,\quad\forall u>0.
\end{equation}
Similarly, by substituting $(u,m)=(1,1),(2,1),(3,1)\cdots$, one gets
$F(u,2)=0,\forall u>0$. In general, we have
\begin{equation}
F(u,m)=0,\quad\forall u>0.
\end{equation}
Finally we substitute $u=0$ in the Eq.(\ref{recurrence}) to obtain%
\begin{equation}
-(1+2m)F(0,m)+2mF(1,m)+F(0,m+1)=0,
\end{equation}
which implies%
\begin{equation}
F(m+1)=(2m+1)F(m).
\end{equation}
Eq.(\ref{lead}) is thus proved by mathematical induction. Note that the case
for $L=0$ corresponds to $f_{2}=0$ in the above calculation. We thus have
proved the nontrivial part of Eq.(\ref{4.3}) for $L=0,1.$

For $L=1,$ the vanishing of the coefficients of $(-\tilde{t})^{0},(-\tilde
{t})^{-1},...,(-\tilde{t})^{-m+1}$ terms on the LHS of Eq.(\ref{4.3}) means,
for $1\leqslant i\leqslant m$,
\begin{align}
G(m,i)  &  \equiv\sum_{j=0}^{m+i}(-1)^{j-i}{\binom{2m}{j+m-i}}\left[
s(j+m-1-i,j)+s(j+m-1-i,j-1)\right] \nonumber\\
&  =0. \label{19}%
\end{align}
Note that for the case of $L=0,$ the second term of Eq.(\ref{19}) vanishes. To
prove the identity Eq.(\ref{19}), we need the recurrence relation of $G(m,i)$
\cite{MK}
\begin{align}
-  &  2(1+m)^{2}(1+2m)G(m,i)+(2+7m+4m^{2})G(m+1,i)\nonumber\\
-  &  2m(1+m)(1+2m)G(m+1,i+1)-mG(m+2,i)=0. \label{20}%
\end{align}
Putting $i=0,1,2..$, and using the fact we have just proved, i.e.
$G(m+1,0)=(2m+1)G(m,0)$, one can show that
\begin{equation}
G(m,i)=0\quad\text{for $1\leqslant{i}\leqslant{m.}$} \label{21}%
\end{equation}
Eq.(\ref{4.3}) is finally proved for the case of $L=0,1$.

\bigskip We now proceed to prove Eq.(\ref{4.3}) for $L=2,3,4,....$To do so, we
rewrite the nontrivial leading term of Eq.(\ref{4.2}) in another form as%
\begin{align}
&  \sum_{j=0}^{2m}(-2m)_{j}\left(  -L-\frac{t}{2}\right)  _{j}\left(
-\frac{2}{t}\right)  ^{j}\frac{1}{j!}\nonumber\\
&  =\sum_{j=0}^{2m}(-1)^{j}{\binom{2m}{j}}\sum_{l=0}^{j}{\binom{j}{l}}\left(
-L\right)  _{j-l}\left(  -\frac{t}{2}\right)  _{l}\left(  -\frac{2}{t}\right)
^{j}\nonumber\\
&  =\sum_{j=0}^{2m}(-1)^{j}{\binom{2m}{j}}\sum_{l=0}^{j}{\binom{j}{l}}\left(
-L\right)  _{j-l}\sum_{s=0}^{l}(-1)^{l-s}s(l,s)\left(  -\frac{t}{2}\right)
^{s}\left(  -\frac{2}{t}\right)  ^{j}\nonumber\\
&  =\sum_{j=0}^{2m}(-1)^{j}{\binom{2m}{j}}\sum_{l=0}^{j}{\binom{j}{l}}\left(
-L\right)  _{j-l}\sum_{s=0}^{l}(-1)^{l-s}s(l,s)\left(  -\frac{2}{t}\right)
^{j-s}\nonumber\\
&  \Longrightarrow\sum_{j=0}^{2m}(-1)^{j}{\binom{2m}{j}}\sum_{l=0}^{j}%
{\binom{j}{l}}\left(  -L\right)  _{j-l}(-1)^{l-j+m}s(l,j-m)\left(  -\frac
{2}{t}\right)  ^{m}\nonumber\\
&  =\sum_{j=0}^{m}{\binom{2m}{j+m}}\sum_{l=0}^{j+m}{\binom{j+m}{l}}%
(-1)^{l+m}\left(  -L\right)  _{j+m-l}s(l,j)\left(  -\frac{2}{t}\right)  ^{m}.
\label{22}%
\end{align}
Thus the nontrivial leading order identity of Eq.(\ref{4.3}) can be written
as
\begin{equation}
\mathcal{F}(m,L)\equiv\sum_{j=0}^{m}\left(
\begin{array}
[c]{c}%
2m\\
j+m
\end{array}
\right)  \sum_{l=0}^{j+m}\left(
\begin{array}
[c]{c}%
j+m\\
l
\end{array}
\right)  (-1)^{l+m}s(l,j)(-L)_{j+m-l}=(2m-1)!!, \label{23}%
\end{equation}
which is independent of $L$ ! We will again use mathematical induction to
prove the identity. Firstly, we note that, for $L=0$ and $L=1$, $\mathcal{F}%
(m,L=0)=(2m-1)!!$ and $\mathcal{F}(m,L=1)=(2m-1)!!$ as have been proved
previously, so Eq.(\ref{23}) is true. Secondly, we notice that $\mathcal{F}%
(m,L)$ satisfies the following recurrence relation \cite{MK}%
\begin{align}
2(1+m)(1+2m)\mathcal{F}(m,1-L)-2(1+m)(1+2m)\mathcal{F}(m,2-L)  & \nonumber\\
+(2+2m+N)\mathcal{F}(1+m,-L)-(3+2m+2N)\mathcal{F}(1+m,1-L)  & \nonumber\\
+(1+N)\mathcal{F}(1+m,2-L)  &  =0. \label{24}%
\end{align}
Eq.(\ref{24}) gives a recurrence relation for $\mathcal{F}(m,L)$ with three
consecutive values of $L.$ One thus can solve the equation and get the final
solution $\mathcal{F}(m,L)=(2m-1)!!$. We thus have proved Eq.(\ref{23}) for
any integer $L.$

To complete the proof of Eq.(\ref{4.3}), we need to show the vanishing of the
coefficients of $(-\tilde{t})^{0},(-\tilde{t})^{-1},...,(-\tilde{t})^{-m+1}$
terms on Eq.(\ref{4.2}) for $L=2,3.4...$At this stage, the authors are unable
to do the exact proof for this case. Instead, we give numerical calculation of
Eq.(\ref{4.2}) for some values of $m$. The results support the identity
Eq.(\ref{4.2}) and can be found in the Appendix B. Moreover, the identity
seems to be valid for arbitrary \textit{real }values $L$ not just integer$.$
So we will take Eq.(\ref{4.2}) as an identity in combinatorial theory
predicted by string theory calculations. We thus have shown that high energy
superstring scattering amplitudes $A_{1}^{(N,2m,q)}$ of Eq.(\ref{A}) in the RR
can be used to extract the ratios $T_{1}^{(N,2m,q)}/T_{1}^{(N,0,0)}$ of
Eq.(\ref{2.15}) in the GR by using the Stirling number identities. That is%
\begin{align}
\lim_{t\rightarrow\infty}\frac{A_{1}^{(N,2m,q)}}{A_{1}^{(N,0,0)}}  &
=\lim_{t\rightarrow\infty}\left(  -\frac{1}{2M_{2}}\right)  ^{2m+q}%
2^{2m}(-t)^{m+2q}U\left(  -2m\,,\,\frac{t}{2}-2m+\frac{3}{2}\,,\,\frac{t}%
{2}\right) \nonumber\\
&  =\left(  -\frac{1}{2M_{2}}\right)  ^{q+m}\frac{\left(  2m-1\right)
!!}{\left(  -M_{2}\right)  ^{m}}=\frac{T_{1}^{(N,2m,q)}}{T_{1}^{(N,0,0)}}.
\end{align}

\subsection{Ratios for $\left\vert N+1,2m+1,q\right\rangle \otimes\left\vert
b_{-\frac{1}{2}}^{P}\right\rangle $}

The ratios $A_{2}^{(N+1,2m+1,q)}/A_{1}^{(N,0,0)}$ can be calculated to be%
\begin{align}
\frac{A_{2}^{(N+1,2m+1,q)}}{A_{1}^{(N,0,0)}}  &  =\left(  -\frac{1}{2M_{2}%
}\right)  ^{2m+q+1}\left(  -\tilde{t}\right)  ^{m}\cdot\left[  (1+t)\sum
_{j=0}^{2m+1}{\binom{1+2m}{j}}\left(  \frac{2}{\tilde{t}}\right)  ^{j}\left(
\frac{1}{2}-\frac{t}{2}\right)  _{j}\right. \nonumber\\
&  \left.  -\tilde{t}\sum_{j=0}^{2m+1}{\binom{1+2m}{j}}\left(  \frac{2}%
{\tilde{t}}\right)  ^{j}\left(  -\frac{1}{2}-\frac{t}{2}\right)  _{j}\right]
.
\end{align}
The bracket in the above equation can be simplified by dropping out the
subleading order terms in the calculation, and one obtains%
\begin{align}
&  (1+t)\sum_{j=0}^{2m+1}{\binom{2m+1}{j}}\left(  \frac{2}{\tilde{t}}\right)
^{j}\left(  \frac{1}{2}-\frac{t}{2}\right)  _{j}-\tilde{t}\sum_{j=0}%
^{2m+1}{\binom{2m+1}{j}}\left(  \frac{2}{\tilde{t}}\right)  ^{j}\left(
-\frac{1}{2}-\frac{t}{2}\right)  _{j}\nonumber\\
&  =(1+t)\sum_{j=0}^{2m+1}(-2m-1)_{j}\left(  -N-\frac{\tilde{t}}{2}\right)
_{j}\frac{(-2/\tilde{t})^{j}}{j!}\nonumber\\
&  -\tilde{t}\sum_{j=0}^{2m+1}(-2m-1)_{j}\left(  -N-1-\frac{\tilde{t}}%
{2}\right)  _{j}\frac{(-2/\tilde{t})^{j}}{j!}\nonumber\\
&  \approx\tilde{t}\sum_{j=0}^{2m+1}(-2m-1)_{j}\left(  -N-\frac{\tilde{t}}%
{2}\right)  _{j}\frac{(-2/\tilde{t})^{j}}{j!}-\tilde{t}\sum_{j=0}%
^{2m+1}(-2m-1)_{j}\left(  -N-1-\frac{\tilde{t}}{2}\right)  _{j}\frac
{(-2/\tilde{t})^{j}}{j!}\nonumber\\
&  =2\left(  2m+1\right)  \cdot\sum_{j=1}^{2m+1}(-2m)_{j-1}\left(
-N-\frac{\tilde{t}}{2}\right)  _{j-1}\frac{(-2/\tilde{t})^{j-1}}{\left(
j-1\right)  !}\nonumber\\
&  =2\left(  2m+1\right)  \cdot\sum_{j=0}^{2m}(-2m)_{j}\left(  -N-\frac
{\tilde{t}}{2}\right)  _{j}\frac{(-2/\tilde{t})^{j}}{j!}%
\end{align}
where we have dropped out the subleading order terms in the second equality of
the calculation. Finally, the ratios can be calculated to be
\begin{align}
\frac{A_{2}^{(N+1,2m+1,q)}}{A_{1}^{(N,0,0)}}  &  =\left(  -\frac{1}{2M_{2}%
}\right)  ^{2m+q+1}\left(  -\tilde{t}\right)  ^{m}\cdot\left[  (1+t)\sum
_{j=0}^{2m+1}{\binom{1+2m}{j}}\left(  \frac{2}{\tilde{t}}\right)  ^{j}\left(
\frac{1}{2}-\frac{t}{2}\right)  _{j}\right. \nonumber\\
&  \left.  -\tilde{t}\sum_{j=0}^{2m+1}{\binom{1+2m}{j}}\left(  \frac{2}%
{\tilde{t}}\right)  ^{j}\left(  -\frac{1}{2}-\frac{t}{2}\right)  _{j}\right]
\nonumber\\
&  \simeq\left(  -\frac{1}{2M_{2}}\right)  ^{2m+q+1}\left(  -\tilde{t}\right)
^{m}2\left(  2m+1\right)  \sum_{j=0}^{2m}(-2m)_{j}\left(  -N-1-\frac{\tilde
{t}}{2}\right)  _{j}\frac{(-2/\tilde{t})^{j}}{j!}. \label{RatioB}%
\end{align}
By using the identity Eq.(\ref{4.3}), one can show that the leading order
coefficients in Eq.(\ref{RatioB}) can be identified with the ratios calculated
in the GR in Eq.(\ref{2.16}). That is%
\begin{equation}
\lim_{t\rightarrow\infty}\frac{A_{2}^{(N+1,2m+1,q)}}{A_{1}^{(N,0,0)}}%
=\frac{T_{2}^{(N+1,2m+1,q)}}{T_{1}^{(N,0,0)}}.
\end{equation}
In the calculation for this case, it is crucial to reduce the upper limit of
the summation $2m+1$ to $2m$. Otherwise, the identity Eq.(\ref{4.3}) will not
be applicable. It is remarkable to see that the leading order coefficients of
Eq.(\ref{RatioB}) can be identified with ratios of Eq.(\ref{2.16}) in the GR.

\subsection{Ratios for $\left\vert N+1,2m,q\right\rangle \otimes\left\vert
b_{-\frac{1}{2}}^{T}\right\rangle $}

The ratios $A_{3}^{(N+1,2m,q)}/A_{1}^{(N,0,0)}$ can be calculated to be%
\begin{equation}
\frac{A_{3}^{(N+1,2m,q)}}{A_{1}^{(N,0,0)}}=\frac{1}{2}\left(  -\frac{1}%
{2M_{2}}\right)  ^{2m+q-1}(-\tilde{t})^{m}\sum_{j=0}^{2m}(-2m)_{j}\left(
-N-1-\frac{\tilde{t}}{2}\right)  _{j}\frac{(-2/\tilde{t})^{j}}{j!}.
\label{RatioC1}%
\end{equation}
By using the identity Eq.(\ref{4.3}), one can show that the leading order
coefficients in Eq.(\ref{RatioC1}) can be identified with the ratios
calculated in the GR in Eq.(\ref{2.17}). That is%
\begin{equation}
\lim_{t\rightarrow\infty}\frac{A_{3}^{(N+1,2m,q)}}{A_{1}^{(N,0,0)}}%
=\frac{T_{3}^{(N+1,2m,q)}}{T_{1}^{(N,0,0)}}.
\end{equation}

\subsection{Ratios for $\left\vert N-1,2m,q-1\right\rangle \otimes\left\vert
b_{-\frac{1}{2}}^{T}b_{-\frac{1}{2}}^{P}b_{-\frac{3}{2}}^{P}\right\rangle $}

The ratios $A_{4}^{(N-1,2m,q-1)}/A_{1}^{(N,0,0)}$ can be calculated to be%
\begin{equation}
\frac{A_{4}^{(N+1,2m+1,q)}}{A_{1}^{(N,0,0)}}=\left(  -\frac{1}{2M_{2}}\right)
^{2m+q}(-\tilde{t})^{m}\sum_{j=0}^{2m}(-2m)_{j}\left(  -N-1-\frac{\tilde{t}%
}{2}\right)  _{j}\frac{(-2/\tilde{t})^{j}}{j!}. \label{RatioD1}%
\end{equation}
By using the identity Eq.(\ref{4.3}), one can show that the leading order
coefficients in Eq.(\ref{RatioD1}) can be identified with the ratios
calculated in the GR in Eq.(\ref{2.18}). That is%
\begin{equation}
\lim_{t\rightarrow\infty}\frac{A_{4}^{(N+1,2m+1,q)}}{A_{1}^{(N,0,0)}}%
=\frac{T_{4}^{(N+1,2m+1,q)}}{T_{1}^{(N,0,0)}}.
\end{equation}

We thus have succeeded in extracting the Ratios of high energy superstring
scattering amplitudes in the GR from the high energy superstring scattering
amplitudes in the RR. In the next section, we will study the subleading order amplitudes.

%

\setcounter{equation}{0}
\renewcommand{\theequation}{\arabic{section}.\arabic{equation}}%

\section{Subleading Order Amplitudes}

In this section, we calculate the next few subleading order amplitudes in the
RR for the mass levels $M_{2}^{2}=2(N+1)=4,6,8.$ Some results for the bosonic
string calculation were presented in Eq.(\ref{22-4}) to Eq.(\ref{25-4}) in the
last section. The relevant kinematic can be found in the Appendix A. We will
see that the ratios derived in section IV persist to subleading order
amplitudes in the RR. For the even mass levels with $(N+1)=\frac{M_{2}^{2}}%
{2}$= odd, we conjecture and give evidences that the existence of these ratios
in the RR persists to all orders in the Regge expansion of all high energy
string scattering amplitudes . For the odd mass levels with $(N+1)=\frac
{M_{2}^{2}}{2}$= even, the existence of these ratios will show up only in the
first $\frac{N+1}{2}+1$ terms in the Regge expansion of the amplitudes. For
the mass level $M_{2}^{2}=4$, there are three states for Eq.(\ref{2.9}), and
we obtain the subleading order expansions as follows.
\begin{align}
|2,0,0\rangle|b_{-\frac{1}{2}}^{T}\rangle &  \rightarrow(\frac{1}{4}%
t^{2}-\frac{1}{4}t)s+(\frac{1}{4}t^{3}+\frac{9}{4}t^{2}+\frac{7}{4}t-\frac
{5}{4})s^{0}\nonumber\\
&  +(\frac{5}{2}t^{3}+18t^{2}+\frac{39}{2}t+4)s^{-1}+O[s^{-2}],
\end{align}%
\begin{align}
|2,2,0\rangle|b_{-\frac{1}{2}}^{T}\rangle &  \rightarrow(\frac{1}{32}%
t^{2}+\frac{1}{8}t+\frac{19}{32})s+(\frac{1}{32}t^{3}+\frac{23}{32}t^{2}%
+\frac{35}{32}t-\frac{19}{32})s^{0}\nonumber\\
&  +(\frac{3}{4}t^{3}-\frac{13}{4}t^{2}-\frac{39}{4}t-\frac{23}{4}%
)s^{-1}+O[s^{-2}],\\
|2,0,1\rangle|b_{-\frac{1}{2}}^{T}\rangle &  \rightarrow(-\frac{1}{16}%
t^{2}-\frac{1}{4}t+\frac{5}{16})s+(-\frac{1}{16}t^{3}-\frac{15}{16}t^{2}%
-\frac{27}{16}t-\frac{29}{16})s^{0}\nonumber\\
&  +(-\frac{3}{4}t^{3}-\frac{17}{4}t^{2}-\frac{45}{4}t-\frac{31}{4}%
)s^{-1}+O[s^{-2}].
\end{align}
In order to simply the notation in the above equations, we have only shown the
second state of the four-point functionss in the correction functions to
represent the scattering amplitudes on the left hand side of each equation. We
find that the ratios of the leading order coefficients of $st^{2}$ are
$\frac{1}{4}:\frac{1}{32}:-\frac{1}{16}$, and it is easy to check that these
are the same as the ratios in the fixed angle limit. Moreover, the ratios
persist in the second subleading order terms $s^{0}t^{3}$ as $\frac{1}%
{4}:\frac{1}{32}:-\frac{1}{16}$. The ratios terminate to this order. We can
also compare the ratios among different worldsheet fermionic states but with
the same mass level $M_{2}^{2}=4$. We have the expansions:
\begin{align}
|2,1,0\rangle|b_{-\frac{1}{2}}^{L}\rangle &  \rightarrow(\frac{1}{16}%
t^{2}-\frac{7}{16}t)s+(\frac{1}{16}t^{3}-\frac{29}{16}t^{2}-\frac{49}%
{16}t-\frac{35}{16})s^{0}\nonumber\\
&  +(-\frac{7}{4}t^{3}-\frac{67}{4}t^{2}-\frac{117}{4}t-\frac{57}{4}%
)s^{-1}+O[s^{-2}],\\
|1,0,0\rangle|b_{-\frac{3}{2}}^{L}\rangle &  \rightarrow(-\frac{1}{8}%
t^{2}-\frac{5}{8}t)s+(-\frac{1}{8}t^{3}-\frac{17}{8}t^{2}-\frac{33}{8}%
t-\frac{25}{8})s^{0}\nonumber\\
&  +(-\frac{7}{4}t^{3}-\frac{61}{4}t^{2}-\frac{109}{4}t-\frac{55}{4}%
)s^{-1}+O[s^{-2}],\\
|0,0,0\rangle|b_{-\frac{1}{2}}^{T}b_{-\frac{1}{2}}^{L}b_{-\frac{3}{2}}%
^{L}\rangle &  \rightarrow(\frac{1}{32}t^{2}+\frac{3}{16}t+\frac{5}%
{32})s+(\frac{1}{32}t^{3}+\frac{15}{32}t^{2}+\frac{27}{32}t+\frac{13}%
{32})s^{0}\nonumber\\
&  +(\frac{1}{2}t^{3}+\frac{7}{2}t^{2}+\frac{11}{2}t+\frac{5}{2}%
)s^{-1}+O[s^{-2}].
\end{align}
The ratios of the leading order coefficients are proportional to that of state
$|2,0,0\rangle|b_{-\frac{1}{2}}^{T}\rangle,$ and can be calculated to be
\begin{equation}
|2,0,0\rangle|b_{-\frac{1}{2}}^{T}\rangle:|2,1,0\rangle|b_{-\frac{1}{2}}%
^{L}\rangle:|1,0,0\rangle|b_{-\frac{3}{2}}^{L}\rangle:|0,0,0\rangle
|b_{-\frac{1}{2}}^{T}b_{-\frac{1}{2}}^{L}b_{-\frac{3}{2}}^{L}\rangle=\frac
{1}{4}:\frac{1}{16}:-\frac{1}{8}:\frac{1}{32}.
\end{equation}
They again match with the ratios in the fixed angle limit. One can also find
that the second subleading order ratios are the same $\frac{1}{4}:\frac{1}%
{16}:-\frac{1}{8}:\frac{1}{32}$. Again the ratios terminate to this order.

For the mass level $M_{2}^{2}=6$, there are three states in Eq.(\ref{2.9}). We
again calculate the subleading order expansions. Interestingly, in this case
the ratios of the coefficients seem to be the same in all orders as can be
seen in the following:
\begin{align}
|3,0,0\rangle|b_{-\frac{1}{2}}^{T}\rangle &  \rightarrow\sqrt{-t}(\frac{1}%
{8}t^{2}-\frac{1}{8}t)s^{2}+\sqrt{-t}(\frac{3}{16}t^{3}+\frac{25}{16}%
t^{2}+\frac{25}{16}t-\frac{21}{16})s\nonumber\\
&  +\sqrt{-t}(\frac{3}{64}t^{4}+\frac{197}{64}t^{3}+\frac{625}{32}t^{2}%
+\frac{743}{32}t+\frac{411}{64})s^{0}+O[s^{-1}],\\
|3,2,0\rangle|b_{-\frac{1}{2}}^{T}\rangle &  \rightarrow\sqrt{-t}(\frac{1}%
{96}t^{2}-\frac{1}{48}t+\frac{11}{32})s^{2}+\sqrt{-t}(\frac{1}{64}t^{3}%
+\frac{13}{32}t-\frac{5}{8})s\nonumber\\
&  +\sqrt{-t}(\frac{1}{256}t^{4}+\frac{9}{128}t^{3}-\frac{925}{256}t^{2}%
-\frac{729}{64}t-\frac{1481}{256})s^{0}+O[s^{-1}],
\end{align}%
\begin{align}
|3,0,1\rangle|b_{-\frac{1}{2}}^{T}\rangle &  \rightarrow\sqrt{-t}(-\frac
{1}{16\sqrt{6}}t^{2}-\frac{3}{8\sqrt{6}}t+\frac{7}{16\sqrt{6}})s^{2}%
\nonumber\\
&  +\sqrt{-t}(-\frac{3}{32\sqrt{6}}t^{3}-\frac{3}{2\sqrt{6}}t^{2}-\frac
{51}{16\sqrt{6}}t-\frac{19}{4\sqrt{6}})s\nonumber\\
&  +\sqrt{-t}(-\frac{3}{128\sqrt{6}}t^{4}-\frac{111}{64\sqrt{6}}t^{3}%
-\frac{1841}{128\sqrt{6}}t^{2}-\frac{1209}{32\sqrt{6}}t-\frac{3573}%
{128\sqrt{6}})s^{0}+O[s^{-1}].
\end{align}
We find that the ratios of the leading order coefficients of $s^{2}t^{5/2}$
are $\frac{1}{8}:\frac{1}{96}:-\frac{1}{16\sqrt{6}}$, and they agree with the
ratios in the fixed angle limit. The ratios of the second and the third order
coefficients of $st^{7/2}$ and $s^{0}t^{9/2}$ are $\frac{3}{16}:\frac{1}%
{64}:-\frac{3}{32\sqrt{6}}$ and $\frac{3}{64}:\frac{1}{256}:-\frac{3}%
{128\sqrt{6}}$, respectively. We find that these two set of ratios are the
same with one another. We predict that the ratios persist to all orders in the expansions.

The expansions among different worldsheet fermionic states but with same mass
level $M_{2}^{2}=6$ are
\begin{align}
|3,0,0\rangle|b_{-\frac{1}{2}}^{L}\rangle &  \rightarrow\sqrt{-t}(\frac{1}%
{48}t^{2}-\frac{17}{48}t)s^{2}+\sqrt{-t}(\frac{1}{32}t^{3}-\frac{151}{96}%
t^{2}-\frac{295}{96}t-\frac{119}{32})s\nonumber\\
&  +\sqrt{-t}(\frac{1}{128}t^{4}-\frac{249}{128}t^{3}-\frac{1317}{64}%
t^{2}-\frac{2883}{64}t-\frac{3831}{128})s^{0}+O[s^{-1}],\\
|2,0,0\rangle|b_{-\frac{3}{2}}^{L}\rangle &  \rightarrow\sqrt{-t}(-\frac
{1}{8\sqrt{6}}t^{2}-\frac{7}{8\sqrt{6}}t)s^{2}\nonumber\\
&  +\sqrt{-t}(-\frac{3}{16\sqrt{6}}t^{3}-\frac{57}{16\sqrt{6}}t^{2}-\frac
{129}{16\sqrt{6}}t-\frac{147}{16\sqrt{6}})s\nonumber\\
&  +\sqrt{-t}(-\frac{3}{64\sqrt{6}}t^{4}-\frac{285}{64\sqrt{6}}t^{3}%
-\frac{1289}{32\sqrt{6}}t^{2}-\frac{2831}{32\sqrt{6}}t-\frac{4011}{64\sqrt{6}%
})s^{0}+O[s^{-1}],\\
|1,0,0\rangle|b_{-\frac{1}{2}}^{T}b_{-\frac{1}{2}}^{L}b_{-\frac{3}{2}}%
^{L}\rangle &  \rightarrow\sqrt{-t}(\frac{1}{96}t^{2}+\frac{1}{12}t+\frac
{7}{96})s^{2}+\sqrt{-t}(\frac{1}{64}t^{3}+\frac{9}{32}t^{2}+\frac{31}%
{48}t+\frac{61}{96})s\nonumber\\
&  +\sqrt{-t}(\frac{1}{256}t^{4}+\frac{77}{192}t^{3}+\frac{2531}{768}%
t^{2}+\frac{643}{96}t+\frac{3569}{768})s^{0}+O[s^{-1}].
\end{align}
The ratios of the leading order coefficients are given by
\begin{align}
|3,0,0\rangle|b_{-\frac{1}{2}}^{T}\rangle &  :|3,1,0\rangle|b_{-\frac{1}{2}%
}^{L}\rangle:|2,0,0\rangle|b_{-\frac{3}{2}}^{L}\rangle:|1,0,0\rangle
|b_{-\frac{1}{2}}^{T}b_{-\frac{1}{2}}^{L}b_{-\frac{3}{2}}^{L}\rangle
\nonumber\\
&  =\frac{1}{8}:\frac{1}{48}:-\frac{1}{8\sqrt{6}}:\frac{1}{96}.
\end{align}
We have checked that they agree with the ratios in the fixed angle limit. The
second and the third subleading order ratios are $\frac{3}{16}:\frac{1}%
{32}:-\frac{3}{16\sqrt{6}}:\frac{1}{64}$ and $\frac{3}{64}:\frac{1}%
{128}:-\frac{3}{64\sqrt{6}}:\frac{1}{256},$ respectively. Again they agree
with the ratios in the fixed angle limit. We expect that the ratios persist to
all orders in the expansions.

For the mass level $M_{2}^{2}=8$, there are six states in Eq.(\ref{2.9})%
\begin{align}
|4,0,0\rangle|b_{-\frac{1}{2}}^{T}\rangle &  \rightarrow\sqrt{-t}(\frac{1}%
{16}t^{3}-\frac{1}{16}t^{2})s^{3}+\sqrt{-t}(\frac{1}{8}t^{4}+t^{3}+\frac{5}%
{4}t^{2}-\frac{9}{8}t)s^{2}\nonumber\\
&  +\sqrt{-t}(\frac{1}{16}t^{5}+\frac{45}{16}t^{4}+\frac{139}{8}t^{3}%
+\frac{91}{4}t^{2}+\frac{137}{16}t-\frac{81}{16})s^{1}\nonumber\\
&  +\sqrt{-t}(\frac{7}{4}t^{5}+\frac{193}{4}t^{4}+\frac{2013}{8}t^{3}%
+\frac{2899}{8}t^{2}+\frac{1381}{8}t+\frac{171}{8})s^{0}+O[s^{-1}],\\
|4,2,0\rangle|b_{-\frac{1}{2}}^{T}\rangle &  \rightarrow\sqrt{-t}(\frac
{1}{256}t^{3}-\frac{3}{64}t^{2}+\frac{51}{256}t)s^{3}\nonumber\\
&  +\sqrt{-t}(\frac{1}{128}t^{4}-\frac{49}{256}t^{3}+\frac{7}{256}t^{2}%
-\frac{251}{256}t+\frac{459}{256})s^{2}\nonumber\\
&  +\sqrt{-t}(\frac{1}{256}t^{5}-\frac{63}{256}t^{4}-\frac{9}{2}t^{3}%
-\frac{893}{64}t^{2}-\frac{1465}{256}t-\frac{1797}{256})s^{1}\nonumber\\
&  +\sqrt{-t}(-\frac{13}{128}t^{5}-\frac{1273}{128}t^{4}-\frac{6419}{64}%
t^{3}-\frac{15167}{64}t^{2}-\frac{26093}{128}t-\frac{5801}{128})s^{0}%
\nonumber\\
&  +O[s^{-1}],\\
|4,4,0\rangle|b_{-\frac{1}{2}}^{T}\rangle &  \rightarrow\sqrt{-t}(\frac
{3}{4096}t^{3}-\frac{129}{4096}t^{2}-\frac{979}{4096}t-\frac{2895}{4096}%
)s^{3}\nonumber\\
&  +\sqrt{-t}(\frac{3}{2048}t^{4}-\frac{327}{2048}t^{3}-\frac{5899}{2048}%
t^{2}-\frac{8917}{2048}t+\frac{2235}{512})s^{2}\nonumber\\
&  +\sqrt{-t}(\frac{3}{4096}t^{5}-\frac{1017}{4096}t^{4}-\frac{23853}%
{2048}t^{3}-\frac{18573}{2048}t^{2}+\frac{76519}{4096}t-\frac{23133}%
{4096})s^{1}\nonumber\\
&  +\sqrt{-t}(-\frac{123}{1024}t^{5}-\frac{19031}{1024}t^{4}-\frac{24895}%
{512}t^{3}+\frac{10657}{512}t^{2}+\frac{109321}{1024}t+\frac{81701}%
{1024})s^{0}\nonumber\\
&  +O[s^{-1}],\\
|4,0,1\rangle|b_{-\frac{1}{2}}^{T}\rangle &  \rightarrow\sqrt{-t}(-\frac
{1}{64\sqrt{2}}t^{3}-\frac{1}{8\sqrt{2}}t^{2}+\frac{9}{64\sqrt{2}}%
t)s^{3}\nonumber\\
&  +\sqrt{-t}(-\frac{1}{32\sqrt{2}}t^{4}-\frac{35}{64\sqrt{2}}t^{3}-\frac
{83}{64\sqrt{2}}t^{2}-\frac{161}{64\sqrt{2}}t+\frac{81}{64\sqrt{2}}%
)s^{2}\nonumber\\
&  +\sqrt{-t}(-\frac{1}{64\sqrt{2}}t^{5}-\frac{53}{64\sqrt{2}}t^{4}-\frac
{125}{16\sqrt{2}}t^{3}-\frac{181}{8\sqrt{2}}t^{2}-\frac{1187}{64\sqrt{2}%
}t-\frac{891}{64\sqrt{2}})s^{1}\nonumber\\
&  +\sqrt{-t}(-\frac{13}{32\sqrt{2}}t^{5}-\frac{401}{32\sqrt{2}}t^{4}%
-\frac{1665}{16\sqrt{2}}t^{3}-\frac{4397}{16\sqrt{2}}t^{2}-\frac{8401}%
{32\sqrt{2}}t-\frac{2165}{32\sqrt{2}})s^{0}\nonumber\\
&  +O[s^{-1}],
\end{align}%
\begin{align}
|4,0,2\rangle|b_{-\frac{1}{2}}^{T}\rangle &  \rightarrow\sqrt{-t}(\frac
{1}{512}t^{3}+\frac{17}{512}t^{2}+\frac{63}{512}t-\frac{81}{512}%
)s^{3}\nonumber\\
&  +\sqrt{-t}(\frac{1}{256}t^{4}+\frac{27}{256}t^{3}+\frac{163}{256}%
t^{2}+\frac{89}{256}t+\frac{45}{16})s^{2}\nonumber\\
&  +\sqrt{-t}(\frac{1}{512}t^{5}+\frac{53}{512}t^{4}+\frac{365}{256}%
t^{3}+\frac{1413}{256}t^{2}+\frac{5685}{512}t-\frac{2095}{512})s^{1}%
\nonumber\\
&  +\sqrt{-t}(\frac{1}{32}t^{5}+\frac{29}{32}t^{4}+\frac{547}{32}t^{3}%
+\frac{1893}{32}t^{2}+90t+\frac{945}{16})s^{0}+O[s^{-1}],\\
|4,2,1\rangle|b_{-\frac{1}{2}}^{T}\rangle &  \rightarrow\sqrt{-t}(-\frac
{1}{1024\sqrt{2}}t^{3}+\frac{3}{1024\sqrt{2}}t^{2}+\frac{57}{1024\sqrt{2}%
}t-\frac{459}{1024\sqrt{2}})s^{3}\nonumber\\
&  +\sqrt{-t}(-\frac{1}{512\sqrt{2}}t^{4}+\frac{33}{512\sqrt{2}}t^{3}%
+\frac{421}{512\sqrt{2}}t^{2}-\frac{309}{512\sqrt{2}}t+\frac{297}{64\sqrt{2}%
})s^{2}\nonumber\\
&  +\sqrt{-t}(-\frac{1}{1024\sqrt{2}}t^{5}+\frac{159}{1024\sqrt{2}}t^{4}%
+\frac{2139}{512\sqrt{2}}t^{3}\nonumber\\
&  +\frac{3983}{512\sqrt{2}}t^{2}+\frac{27419}{1024\sqrt{2}}t+\frac
{1043}{1024\sqrt{2}})s^{1}\nonumber\\
&  +\sqrt{-t}(\frac{3}{32\sqrt{2}}t^{5}+\frac{1073}{128\sqrt{2}}t^{4}%
+\frac{1203}{32\sqrt{2}}t^{3}+\frac{11931}{64\sqrt{2}}t^{2}+\frac{1913}%
{8\sqrt{2}}t+\frac{13569}{128\sqrt{2}})s^{0}\nonumber\\
&  +O[s^{-1}].
\end{align}
We find that the ratios of the leading order coefficients of $s^{3}t^{3}$ are
$\frac{1}{16}:\frac{1}{256}:\frac{3}{4096}:-\frac{1}{64\sqrt{2}}:\frac{1}%
{512}:-\frac{1}{1024\sqrt{2}}$, and they agree with the ratios in the fixed
angle limit. The ratios of the second and the third order coefficients of
$s^{2}t^{4}$ and $s^{1}t^{5}$ are $\frac{1}{8}:\frac{1}{128}:\frac{3}%
{2048}:-\frac{1}{32\sqrt{2}}:\frac{1}{256}:-\frac{1}{512\sqrt{2}}$ and
$\frac{1}{16}:\frac{1}{256}:\frac{3}{4096}:-\frac{1}{64\sqrt{2}}:\frac{1}%
{512}:-\frac{1}{1024\sqrt{2}}$, respectively. We find that the above three
ratios are the same with one another. One can see that the ratios terminate at
the order $s^{0}t^{5}$ as expected.

The expansions among different worldsheet fermionic states but with the same
mass level $M_{2}^{2}=8$ are
\begin{align}
|4,0,0\rangle|b_{-\frac{1}{2}}^{L}\rangle &  \rightarrow\sqrt{-t}(\frac
{1}{128}t^{3}-\frac{31}{128}t^{2})s^{3}+\sqrt{-t}(\frac{1}{64}t^{4}-\frac
{19}{16}t^{3}-\frac{41}{16}t^{2}-\frac{279}{64}t)s^{2}\nonumber\\
&  +\sqrt{-t}(\frac{1}{128}t^{5}-\frac{231}{128}t^{4}-\frac{1289}{64}%
t^{3}-\frac{3385}{64}t^{2}-\frac{5527}{128}t-\frac{2511}{128})s^{1}\nonumber\\
&  +\sqrt{-t}(-\frac{55}{64}t^{5}-\frac{2731}{64}t^{4}-\frac{9975}{32}%
t^{3}-\frac{22423}{32}t^{2}-\frac{39563}{64}t-\frac{11223}{64})s^{0}%
\nonumber\\
&  +O[s^{-1}],\\
|4,2,0\rangle|b_{-\frac{1}{2}}^{L}\rangle &  \rightarrow\sqrt{-t}(\frac
{3}{2048}t^{3}-\frac{131}{1024}t^{2}+\frac{759}{2048}t)s^{3}\nonumber\\
&  +\sqrt{-t}(\frac{3}{1024}t^{4}-\frac{1355}{2048}t^{3}+\frac{7207}%
{2048}t^{2}+\frac{3111}{2048}t+\frac{6831}{2048}t)s^{2}\nonumber\\
&  +\sqrt{-t}(\frac{3}{2048}t^{5}-\frac{2153}{2048}t^{4}+\frac{8169}%
{1024}t^{3}+\frac{32837}{1024}t^{2}+\frac{133867}{2048}t+\frac{41631}%
{2048})s^{1}\nonumber\\
&  +\sqrt{-t}(-\frac{265}{512}t^{5}+\frac{2259}{512}t^{4}+\frac{37975}%
{256}t^{3}+\frac{143935}{256}t^{2}+\frac{342379}{512}t+\frac{140223}%
{512})s^{0}\nonumber\\
&  +O[s^{-1}],
\end{align}%
\begin{align}
|4,0,1\rangle|b_{-\frac{1}{2}}^{L}\rangle &  \rightarrow\sqrt{-t}(-\frac
{1}{512\sqrt{2}}t^{3}+\frac{11}{256\sqrt{2}}t^{2}+\frac{279}{512\sqrt{2}%
}t)s^{3}\nonumber\\
&  +\sqrt{-t}(-\frac{1}{256\sqrt{2}}t^{4}+\frac{181}{512\sqrt{2}}t^{3}%
+\frac{1687}{512\sqrt{2}}t^{2}+\frac{943}{512\sqrt{2}}t+\frac{2511}%
{512\sqrt{2}})s^{2}\nonumber\\
&  +\sqrt{-t}(-\frac{1}{512\sqrt{2}}t^{5}+\frac{363}{512\sqrt{2}}t^{4}%
+\frac{2081}{256\sqrt{2}}t^{3}+\frac{8293}{256\sqrt{2}}t^{2}+\frac
{27967}{512\sqrt{2}}t+\frac{3915}{512\sqrt{2}})s^{1}\nonumber\\
&  +\sqrt{-t}(\frac{51}{128\sqrt{2}}t^{5}+\frac{1467}{128\sqrt{2}}t^{4}%
+\frac{7523}{64\sqrt{2}}t^{3}+\frac{25623}{64\sqrt{2}}t^{2}+\frac
{63223}{128\sqrt{2}}t+\frac{28679}{128\sqrt{2}})s^{0}\nonumber\\
&  +O[s^{-1}],\\
|4,2,1\rangle|b_{-\frac{1}{2}}^{L}\rangle &  \rightarrow\sqrt{-t}(-\frac
{3}{8192\sqrt{2}}t^{3}+\frac{235}{8192\sqrt{2}}t^{2}+\frac{1599}{8192\sqrt{2}%
}t-\frac{6831}{8192\sqrt{2}})s^{3}\nonumber\\
&  +\sqrt{-t}(-\frac{3}{4096\sqrt{2}}t^{4}+\frac{803}{4096\sqrt{2}}t^{3}%
-\frac{809}{4096\sqrt{2}}t^{2}-\frac{43879}{4096\sqrt{2}}t+\frac{861}%
{512\sqrt{2}})s^{2}\nonumber\\
&  +\sqrt{-t}(-\frac{3}{8192\sqrt{2}}t^{5}+\frac{3029}{8192\sqrt{2}}%
t^{4}-\frac{21333}{4096\sqrt{2}}t^{3}\nonumber\\
&  -\frac{115237}{4096\sqrt{2}}t^{2}-\frac{150707}{8192\sqrt{2}}%
t-\frac{326379}{8192\sqrt{2}})s^{1}\nonumber\\
&  +\sqrt{-t}(\frac{829}{4096\sqrt{2}}t^{5}-\frac{45011}{4096\sqrt{2}}%
t^{4}-\frac{68377}{2048\sqrt{2}}t^{3}\nonumber\\
&  -\frac{657673}{2048\sqrt{2}}t^{2}-\frac{2228619}{4096\sqrt{2}}%
t-\frac{807579}{4096\sqrt{2}})s^{0}\nonumber\\
&  +O[s^{-1}],\\
|3,0,0\rangle|b_{-\frac{3}{2}}^{L}\rangle &  \rightarrow\sqrt{-t}(\frac
{1}{32\sqrt{2}}t^{3}+\frac{9}{32\sqrt{2}}t^{2})s^{3}\nonumber\\
&  +\sqrt{-t}(\frac{1}{16\sqrt{2}}t^{4}+\frac{21}{16\sqrt{2}}t^{3}+\frac
{53}{16\sqrt{2}}t^{2}+\frac{81}{16\sqrt{2}}t)s^{2}\nonumber\\
&  +\sqrt{-t}(\frac{1}{32\sqrt{2}}t^{5}+\frac{69}{32\sqrt{2}}t^{4}+\frac
{327}{16\sqrt{2}}t^{3}+\frac{847}{16\sqrt{2}}t^{2}+\frac{1485}{32\sqrt{2}%
}t+\frac{729}{32\sqrt{2}})s^{1}\nonumber\\
&  +\sqrt{-t}(\frac{9}{8\sqrt{2}}t^{5}+\frac{315}{8\sqrt{2}}t^{4}+\frac
{284}{\sqrt{2}}t^{3}+\frac{658}{\sqrt{2}}t^{2}+\frac{4771}{8\sqrt{2}}%
t+\frac{1377}{8\sqrt{2}})s^{0}+O[s^{-1}],\\
|3,2,0\rangle|b_{-\frac{3}{2}}^{L}\rangle &  \rightarrow\sqrt{-t}(\frac
{1}{512\sqrt{2}}t^{3}-\frac{11}{256\sqrt{2}}t^{2}-\frac{279}{512\sqrt{2}%
}t)s^{3}\nonumber\\
&  +\sqrt{-t}(\frac{1}{256\sqrt{2}}t^{4}-\frac{147}{512\sqrt{2}}t^{3}%
-\frac{2187}{512\sqrt{2}}t^{2}-\frac{1477}{512\sqrt{2}}t-\frac{2511}%
{512\sqrt{2}})s^{2}\nonumber\\
&  +\sqrt{-t}(\frac{1}{512\sqrt{2}}t^{5}-\frac{267}{512\sqrt{2}}t^{4}%
-\frac{865}{64\sqrt{2}}t^{3}-\frac{5919}{128\sqrt{2}}t^{2}-\frac
{37009}{512\sqrt{2}}t-\frac{8721}{512\sqrt{2}})s^{1}\nonumber\\
&  +\sqrt{-t}(-\frac{71}{256\sqrt{2}}t^{5}-\frac{5143}{256\sqrt{2}}t^{4}%
-\frac{26457}{128\sqrt{2}}t^{3}\nonumber\\
&  -\frac{81497}{128\sqrt{2}}t^{2}-\frac{184135}{256\sqrt{2}}t-\frac
{75127}{256\sqrt{2}})s^{0}\nonumber\\
&  +O[s^{-1}],
\end{align}%
\begin{align}
|3,0,1\rangle|b_{-\frac{3}{2}}^{L}\rangle &  \rightarrow\sqrt{-t}(-\frac
{1}{256}t^{3}-\frac{9}{128}t^{2}-\frac{81}{256}t^{2})s^{3}\nonumber\\
&  +\sqrt{-t}(-\frac{1}{128}t^{4}-\frac{57}{256}t^{3}-\frac{373}{256}%
t^{2}-\frac{279}{256}t-\frac{729}{256})s^{2}\nonumber\\
&  +\sqrt{-t}(-\frac{1}{256}t^{5}-\frac{61}{256}t^{4}-\frac{221}{64}%
t^{3}-\frac{465}{32}t^{2}-\frac{5955}{256}t-\frac{243}{256})s^{1}\nonumber\\
&  +\sqrt{-t}(-\frac{11}{128}t^{5}-\frac{447}{128}t^{4}-\frac{2847}{64}%
t^{3}-\frac{10587}{64}t^{2}-\frac{27511}{128}t-\frac{13131}{128}%
)s^{0}\nonumber\\
&  +O[s^{-1}],
\end{align}%
\begin{align}
|2,0,0\rangle|b_{-\frac{1}{2}}^{T}b_{-\frac{1}{2}}^{L}b_{-\frac{3}{2}}%
^{L}\rangle &  \rightarrow\sqrt{-t}(\frac{1}{256}t^{3}+\frac{5}{128}%
t^{2}+\frac{9}{256}t)s^{3}\nonumber\\
&  +\sqrt{-t}(\frac{1}{128}t^{4}+\frac{41}{256}t^{3}+\frac{109}{256}%
t^{2}+\frac{151}{256}t+\frac{81}{256})s^{2}\nonumber\\
&  +\sqrt{-t}(\frac{1}{256}t^{5}+\frac{73}{256}t^{4}+\frac{83}{32}t^{3}%
+\frac{409}{64}t^{2}+\frac{1503}{256}t+\frac{459}{256})s^{1}\nonumber\\
&  +\sqrt{-t}(\frac{21}{128}t^{5}+\frac{705}{128}t^{4}+\frac{2263}{64}%
t^{3}+\frac{4987}{64}t^{2}+\frac{9085}{128}t+\frac{2953}{128})s^{0}\nonumber\\
&  +O[s^{-1}],\\
|2,2,0\rangle|b_{-\frac{1}{2}}^{T}b_{-\frac{1}{2}}^{L}b_{-\frac{3}{2}}%
^{L}\rangle &  \rightarrow\sqrt{-t}(\frac{1}{4096}t^{3}-\frac{41}{4096}%
t^{2}-\frac{501}{4096}t-\frac{459}{4096})s^{3}\nonumber\\
&  +\sqrt{-t}(\frac{1}{2048}t^{4}-\frac{29}{512}t^{3}-\frac{825}{1024}%
t^{2}-\frac{321}{512}t+\frac{249}{2048})s^{2}\nonumber\\
&  +\sqrt{-t}(\frac{1}{4096}t^{5}-\frac{371}{4096}t^{4}-\frac{4715}{4096}%
t^{3}-\frac{14947}{4096}t^{2}-\frac{39131}{4096}t-\frac{18295}{4096}%
)s^{1}\nonumber\\
&  +\sqrt{-t}(-\frac{45}{1024}t^{5}-\frac{3473}{1024}t^{4}-\frac{17217}%
{1024}t^{3}-\frac{46833}{1024}t^{2}\nonumber\\
&  -\frac{97697}{1024}t-\frac{35037}{1024})s^{0}+O[s^{-1}],\\
|2,0,1\rangle|b_{-\frac{1}{2}}^{T}b_{-\frac{1}{2}}^{L}b_{-\frac{3}{2}}%
^{L}\rangle &  \rightarrow\sqrt{-t}(-\frac{1}{1024\sqrt{2}}t^{3}-\frac
{19}{1024\sqrt{2}}t^{2}-\frac{99}{1024\sqrt{2}}t-\frac{81}{1024\sqrt{2}}%
)s^{3}\nonumber\\
&  +\sqrt{-t}(-\frac{1}{512\sqrt{2}}t^{4}-\frac{13}{256\sqrt{2}}t^{3}%
-\frac{41}{128\sqrt{2}}t^{2}-\frac{47}{256\sqrt{2}}t+\frac{45}{512\sqrt{2}%
})s^{2}\nonumber\\
&  +\sqrt{-t}(-\frac{1}{1024\sqrt{2}}t^{5}-\frac{57}{1024\sqrt{2}}t^{4}%
-\frac{405}{512\sqrt{2}}t^{3}\nonumber\\
&  -\frac{1945}{512\sqrt{2}}t^{2}-\frac{5573}{1024\sqrt{2}}t-\frac
{2437}{1024\sqrt{2}})s^{1}\nonumber\\
&  +\sqrt{-t}(-\frac{3}{128\sqrt{2}}t^{5}-\frac{57}{64\sqrt{2}}t^{4}%
-\frac{737}{64\sqrt{2}}t^{3}\nonumber\\
&  -\frac{615}{16\sqrt{2}}t^{2}-\frac{5859}{128\sqrt{2}}t-\frac{1151}%
{64\sqrt{2}})s^{0}+O[s^{-1}].
\end{align}
The ratios of the leading order coefficients are given by
\begin{align}
|4,0,0\rangle|b_{-\frac{1}{2}}^{T}\rangle &  :|4,0,0\rangle|b_{-\frac{1}{2}%
}^{L}\rangle:|4,2,0\rangle|b_{-\frac{1}{2}}^{L}\rangle:|4,0,1\rangle
|b_{-\frac{1}{2}}^{L}\rangle:|4,2,1\rangle|b_{-\frac{1}{2}}^{L}\rangle
\nonumber\\
&  :|3,0,0\rangle|b_{-\frac{3}{2}}^{L}\rangle:|3,2,0\rangle|b_{-\frac{3}{2}%
}^{L}\rangle:|3,0,1\rangle|b_{-\frac{3}{2}}^{L}\rangle:|2,0,0\rangle
|b_{-\frac{1}{2}}^{T}b_{-\frac{1}{2}}^{L}b_{-\frac{3}{2}}^{L}\rangle
\nonumber\\
&  :|2,2,0\rangle|b_{-\frac{1}{2}}^{T}b_{-\frac{1}{2}}^{L}b_{-\frac{3}{2}}%
^{L}\rangle:|2,0,1\rangle|b_{-\frac{1}{2}}^{T}b_{-\frac{1}{2}}^{L}b_{-\frac
{3}{2}}^{L}\rangle\nonumber\\
&  =\frac{1}{16}:\frac{1}{128}:\frac{3}{2048}:-\frac{1}{512\sqrt{2}}:-\frac
{3}{8192\sqrt{2}}\nonumber\\
&  :\frac{1}{32\sqrt{2}}:\frac{1}{512\sqrt{2}}:-\frac{1}{256}:\frac{1}%
{256}\nonumber\\
&  :\frac{1}{4096}:-\frac{1}{1024\sqrt{2}}.
\end{align}
They agree with the ratios in the fixed angle limit. It can be checked that
the second and the third subleading order coefficients of $s^{2}t^{4}$ and
$s^{1}t^{5}$ have the same ratios. The ratios terminate in the fourth
subleading order $s^{0}t^{5}$ coefficients as expected.

\bigskip%

\setcounter{equation}{0}
\renewcommand{\theequation}{\arabic{section}.\arabic{equation}}%

\section{Conclusion}

In this paper, we calculate high energy massive superstring scattering
amplitudes in the Regge regime (RR). We explicitly calculate four classes of
high energy Regge scattering amplitudes. As an application, we demonstrate the
universal power-law behavior for all massive string scattering amplitudes in
the RR. In particular, the amplitude gives the correct intercept $a_{0}%
=\frac{1}{2}$ of fermonic string theory. These results generalize the well
known results for the case of high energy four-point tachyon scattering
amplitudes. Moreover, as in the bosonic string case considered previously
\cite{bosonic}, these amplitudes can be used to extract ratios among high
energy superstring scattering amplitudes in the fixed angle regime. The
calculation relies on a set of Stirling number identities, which we are able
to give only partial proofs of them. For this reason, we give a numerical
"proof" of the whole identities. Hopefully, the complete mathematical proof of
these identities suggested by string theory calculation can be worked out in
the future.

In addition to the leading order calculation, we also study the subleading
order amplitudes in the Regge regime for the first few mass levels. In
particular, we conjecture and give evidences that the existence of the GR
ratios in the RR persists to all orders in the Regge expansion of all string
amplitudes for the even mass level with $(N+1)=\frac{M_{2}^{2}}{2}$= odd. For
the odd mass levels with $(N+1)=\frac{M_{2}^{2}}{2}$= even, the existence of
the GR ratios shows up only in the first $\frac{N+1}{2}+1$ terms in the Regge
expansion of the amplitudes. It will be an interesting challenge to further
study this subleading order effect.

\begin{acknowledgments}
S. He would like to thank Prof. Mei Huang's warm support. He thanks for
discussion about Computer Algebra with Prof. Fr\'{e}d\'{e}ric Chyzak in France
and Prof. Yu Fu Chen in GUCAS. He also thanks for the hospitality of
Department of Electrophysics of National Chiao-Tung University and National
Center for Theoretical Science at Hsinchu during his visit. The authors
appreciated many helpful correspondences of Dr. Manuel Mkauers at RISC,
Austria. The help of numerical work from Rong-Shing Chang is acknowledged.
Finally, we thank the comment from the referee which helps to clarify the
conformal property of the vertex in Eq.(\ref{pq}). This work is supported in
part by the National Science Council, 50 billions project of Ministry of
Education and National Center for Theoretical Science, Taiwan.
\end{acknowledgments}

\appendix%

\setcounter{equation}{0}
\renewcommand{\theequation}{\thesection.\arabic{equation}}%

\section{Kinematic Relations in the RR}

In this appendix, we list the expressions of the kinematic variables we used
in the evaluation of 4-point functions in this paper. For convenience, we take
the center of momentum frame and choose the momenta of particles 1 and 2 to be
along the $X^{1}$-direction. The high energy scattering plane is defined to be
on the $X^{1}-X^{2}$ plane.\begin{figure}[h]
\label{scattering} \setlength{\unitlength}{3pt}
\par
\begin{center}
\begin{picture}(100,100)(-50,-50)
{\large
\put(45,0){\vector(-1,0){42}} \put(-45,0){\vector(1,0){42}}
\put(2,2){\vector(1,1){30}} \put(-2,-2){\vector(-1,-1){30}}
\put(25,2){$k_1$} \put(-27,2){$k_2$} \put(11,20){$-k_3$}
\put(-24,-15){$-k_4$}
\put(40,0){\vector(0,-1){10}} \put(-40,0){\vector(0,1){10}}
\put(26,26){\vector(-1,1){7}} \put(-26,-26){\vector(1,-1){7}}
\put(36,-16){$e^{T}(1)$} \put(-44,15){$e^{T}(2)$}
\put(15,36){$e^{T}(3)$} \put(-18,-35){$e^{T}(4)$}
\qbezier(10,0)(10,4)(6,6) \put(12,4){$\theta$}
\put(-55,-45){Fig.1 Kinematic variables in the center of mass frame} }
\end{picture}
\end{center}
\end{figure}

The momenta of the four particles are%

\begin{align}
k_{1}  &  =\left(  +\sqrt{p^{2}+M_{1}^{2}},-p,0\right)  ,\\
k_{2}  &  =\left(  +\sqrt{p^{2}+M_{2}^{2}},+p,0\right)  ,\\
k_{3}  &  =\left(  -\sqrt{q^{2}+M_{3}^{2}},-q\cos\theta,-q\sin\theta\right)
,\\
k_{4}  &  =\left(  -\sqrt{q^{2}+M_{4}^{2}},+q\cos\theta,+q\sin\theta\right)
\end{align}

where $p\equiv\left\vert \mathrm{\vec{p}}\right\vert $, $q\equiv\left\vert
\mathrm{\vec{q}}\right\vert $ and $k_{i}^{2}=-M_{i}^{2}$. In the calculation
of the string scattering amplitudes, we use the following formulas%

\begin{align}
-k_{1}\cdot k_{2}  &  =\sqrt{p^{2}+M_{1}^{2}}\cdot\sqrt{p^{2}+M_{2}^{2}}%
+p^{2}=\dfrac{1}{2}\left(  s-M_{1}^{2}-M_{2}^{2}\right)  ,\\
-k_{2}\cdot k_{3}  &  =-\sqrt{p^{2}+M_{2}^{2}}\cdot\sqrt{q^{2}+M_{3}^{2}%
}+pq\cos\theta=\dfrac{1}{2}\left(  t-M_{2}^{2}-M_{3}^{2}\right)  ,\\
-k_{1}\cdot k_{3}  &  =-\sqrt{p^{2}+M_{1}^{2}}\cdot\sqrt{q^{2}+M_{3}^{2}%
}-pq\cos\theta=\dfrac{1}{2}\left(  u-M_{1}^{2}-M_{3}^{2}\right)
\end{align}
where the Mandelstam variables are defined as usual with%

\begin{equation}
s+t+u=\sum_{i}M_{i}^{2}=2N-1.
\end{equation}
The center of mass energy $E$ is defined as%
\begin{equation}
E=\dfrac{1}{2}\left(  \sqrt{p^{2}+M_{1}^{2}}+\sqrt{p^{2}+M_{2}^{2}}\right)
=\dfrac{1}{2}\left(  \sqrt{q^{2}+M_{3}^{2}}+\sqrt{q^{2}+M_{4}^{2}}\right)  .
\end{equation}
We define the polarizations of the string state on the scattering plane as%

\begin{align}
e^{P}  &  =\frac{1}{M_{2}}\left(  \sqrt{p^{2}+M_{2}^{2}},p,0\right)  ,\\
e^{L}  &  =\frac{1}{M_{2}}\left(  p,\sqrt{p^{2}+M_{2}^{2}},0\right)  ,\\
e^{T}  &  =\left(  0,0,1\right)  .
\end{align}
The projections of the momenta on the scattering plane can be calculated as
(here we only list the ones we need for our calculations)%

\begin{align}
e^{P}\cdot k_{1}  &  =-\frac{1}{M_{2}}\left(  \sqrt{p^{2}+M_{1}^{2}}%
\sqrt{p^{2}+M_{2}^{2}}+p^{2}\right)  ,\label{A13}\\
e^{L}\cdot k_{1}  &  =-\frac{p}{M_{2}}\left(  \sqrt{p^{2}+M_{1}^{2}}%
+\sqrt{p^{2}+M_{2}^{2}}\right)  ,\\
e^{T}\cdot k_{1}  &  =0
\end{align}
and%
\begin{align}
e^{P}\cdot k_{3}  &  =\frac{1}{M_{2}}\left(  \sqrt{q^{2}+M_{3}^{2}}\sqrt
{p^{2}+M_{2}^{2}}-pq\cos\theta\right)  ,\\
e^{L}\cdot k_{3}  &  =\frac{1}{M_{2}}\left(  p\sqrt{q^{2}+M_{3}^{2}}%
-q\sqrt{p^{2}+M_{2}^{2}}\cos\theta\right)  ,\\
e^{T}\cdot k_{3}  &  =-q\sin\theta. \label{A18}%
\end{align}
We now expand the kinematic relations to the subleading orders in the RR. We
first express all kinematic variables in terms of $s$ and $t$, and then expand
all relevant quantities in $s:$
\begin{align}
E_{1}  &  =\frac{s-(M_{2}^{2}+2)}{2\sqrt{2}},\\
E_{2}  &  =\frac{s+(M_{2}^{2}+2)}{2\sqrt{2}},\\
|\mathbf{k_{2}}|  &  =\sqrt{E_{1}^{2}+2},\quad|\mathbf{K_{3}}|=\sqrt{\frac
{s}{4}+2};
\end{align}%
\begin{equation}
e_{P}\cdot k_{1}=-\frac{1}{2M_{2}}s+\left(  -\frac{1}{M_{2}}+\frac{M_{2}}%
{2}\right)  ,\quad(\text{exact})
\end{equation}%
\begin{align}
e_{L}\cdot k_{1}  &  =-\frac{1}{2M_{2}}s+\left(  -\frac{1}{M_{2}}+\frac{M_{2}%
}{2}\right)  -{2}M_{2}s^{-1}-2M_{2}(M_{2}^{2}-2)s^{-2}\nonumber\\
&  -2m_{2}(M_{2}^{4}-6M_{2}^{2}+4)s^{-3}-2M_{2}(M_{2}^{6}-12M_{2}^{4}%
+24M_{2}^{2}-8)s^{-4}+O(s^{-5}),
\end{align}%
\begin{equation}
e_{T}\cdot k_{1}=0.
\end{equation}
A key step is to express the scattering angle $\theta$ in terms of $s$ and
$t$. This can be achieved by solving
\begin{equation}
t=-\left(  -(E_{2}-\frac{\sqrt{s}}{2})^{2}+(|\mathbf{k_{2}}|-|\mathbf{k_{3}%
}|\cos\theta)^{2}+|\mathbf{k}_{3}|^{2}\sin^{2}\theta\right)
\end{equation}
to obtain%
\begin{equation}
\theta=\arccos\left(  \frac{s+2t-M_{2}^{2}+6}{\sqrt{s+8}\sqrt{\frac
{(s+2)^{2}-2(s-2)M_{2}^{2}+M_{2}^{4}}{s}}}\right)  .\text{ (exact)}%
\end{equation}
One can then calculate the following expansions which we used in the
subleading order calculation in section V
\begin{equation}
e_{P}\cdot k_{3}=\frac{1}{M_{2}}(E_{2}\frac{\sqrt{s}}{2}-|\mathbf{k_{2}%
}||\mathbf{k_{3}}|\cos\theta)=-\frac{t+2-M_{2}^{2}}{2M_{2}},
\end{equation}%
\begin{align}
e_{L}\cdot k_{3}  &  =\frac{1}{M_{2}}(k_{2}\frac{\sqrt{2}}{2}-E_{2}k_{3}%
\cos\theta)\nonumber\\
&  =-\frac{t+2+M_{2}^{2}}{2M_{2}}-M_{2}ts^{-1}-M_{2}[-4(t+1)+M_{2}%
^{2}(t-2)]s^{-2}\nonumber\\
&  -M_{2}[4(4+3t)-12tM_{2}^{2}+(t-4)M_{2}^{4}]s^{-3}-M_{2}%
[-16(3+2t)+24(2+3t)M_{2}^{2}\nonumber\\
&  -24(-1+t)M_{2}^{4}+(-6+t)M_{2}^{6}]s^{-4}+O(s^{-5}),
\end{align}%
\begin{align}
e_{T}\cdot k_{3}  &  =-|\mathbf{k_{3}}|\sin\theta\nonumber\\
&  =-\sqrt{-t}-\frac{1}{2}\sqrt{-t}(2+t+M_{2}^{2})s^{-1}\nonumber\\
&  \quad-\frac{1}{8\sqrt{-t}}[32+52t+20t^{2}+t^{3}+(32+20t-6t^{2})M_{2}%
^{2}+(8-3t)M_{2}^{4}]s^{-2}\nonumber\\
&  \quad+\frac{1}{16\sqrt{-t}}[320+456t+188t^{2}+22t^{3}+t^{4}%
-(-224+36t+132t^{2}+5t^{3})M_{2}^{2}.\nonumber\\
&  \quad\quad\quad\quad\quad\quad+(-16-122t+15t^{2})M_{2}^{4}+(-24+5t)M_{2}%
^{6}]s^{-3}\nonumber\\
&  \quad+\frac{1}{128(-t)^{3/2}}[1024+12032t+16080t^{2}+7520t^{3}%
+1432t^{4}+136t^{5}+5t^{6}\nonumber\\
&  -4(-512-896t+2232t^{2}+1844t^{3}+170t^{4}+7t^{5})M_{2}^{2}\nonumber\\
&  +2(768-2240t-2372t^{2}+1172t^{3}+35t^{4})M_{2}^{4}\nonumber\\
&  -4(-128+288t-450t^{2}+35t^{3})M_{2}^{6}+(64+240t-35t^{2})M_{2}^{8}%
]s^{-4}+O(s^{-5}).
\end{align}

\section{Numerical Identities}

In this appendix, we give a numerical "proof" of the identity
\begin{align}
&  \sum_{j=0}^{2m}(-2m)_{j}\left(  -L-\frac{\tilde{t}}{2}\right)  _{j}%
\frac{(-2/\tilde{t})^{j}}{j!}\\
&  =0(-\tilde{t})^{0}+0(-\tilde{t})^{-1}+...+0(-\tilde{t})^{-m+1}+\frac
{(2m)!}{m!}(-\tilde{t})^{-m}+\mathit{O}\left\{  \left(  \frac{1}{\tilde{t}%
}\right)  ^{m+1}\right\}  ,
\end{align}
which we used intensively in section IV to rederive the ratios among high
energy scattering amplitudes in the Fixed angle regime from the Regge
scattering amplitudes. The nontrivial identity of Eq.(B.2) has been proved for
arbitrary integers $L$ by using Stirling number identities. However, the "$0$
identities" were exactly proved only for $L=0,1.$ We conjecture that all
identities in Eq.(B.2) are valid for arbitrary \textit{real }$L.$ We have done
the numerical proof of the identity for $m$ up to $m=10.$ Here we give only
results of $m=3$ and $4$
\begin{align}
&  \sum_{j=0}^{6}(-2m)_{j}\left(  a-\frac{\tilde{t}}{2}\right)  _{j}%
\frac{(-2/\tilde{t})^{j}}{j!}\nonumber\\
&  =\frac{120}{(-\tilde{t})^{3}}+\frac{720a^{2}+2640a+2080}{(-\tilde{t})^{4}%
}+\frac{480a^{4}+4160a^{3}+12000a^{2}+12928a+3840}{(-\tilde{t})^{5}%
}\nonumber\\
&  +\frac{64a^{6}+960a^{5}+5440a^{4}+14400a^{3}+17536a^{2}+7680a}{(-\tilde
{t})^{6}},
\end{align}%
\begin{align}
&  \sum_{j=0}^{8}(-2m)_{j}\left(  a-\frac{\tilde{t}}{2}\right)  _{j}%
\frac{(-2/\tilde{t})^{j}}{j!}\nonumber\\
&  =\frac{1680}{(-\tilde{t})^{4}}+\frac{13440a^{2}+67200a+76160}{(-\tilde
{t})^{5}}\nonumber\\
&  +\frac{13440a^{4}+152320a^{3}+595840a^{2}+930048a+467712}{(-\tilde{t})^{6}%
}\\
&  +\frac{3584a^{6}+68096a^{5}+501760a^{4}+1802752a^{3}+3236352a^{2}%
+2608128a+645120}{(-\tilde{t})^{7}}\nonumber\\
&  +\frac{256a^{8}+7168a^{7}+82432a^{6}+501760a^{5}+1732864a^{4}%
+3361792a^{3}+3345408a^{2}+1290240a}{(-\tilde{t})^{8}}%
\end{align}

where $a=-L.$ We can see that $a$ shows up only in the subleading order terms
as expected. For $m=5,$ the nontrivial leading order term is $\frac
{30240}{(-\tilde{t})^{5}}$ as expected. For $m=10,$ the nontrivial leading
order term is $\frac{670442572800}{(-\tilde{t})^{10}}$ as expected.

\end{document}